\documentclass[a4paper,11pt]{article}

\usepackage[T1]{fontenc}
\usepackage[english]{babel}
\usepackage{graphicx,subfigure}
\usepackage{textcomp}
\usepackage{geometry}
\usepackage{bm}
\usepackage{setspace}
\usepackage{color}
\usepackage{latexsym}
\usepackage{amssymb}

\usepackage[numbers]{natbib}

\begin{document}

\begin{center}
\LARGE{Biomimetic bluff body drag reduction by self-adaptive porous flaps.}
\end{center}

\begin{center}
Nicolas Mazellier\footnote{Corresponding author: nicolas.mazellier@univ-orleans.fr} Audrey Feuvrier and Azeddine Kourta
\end{center}


\begin{center}
\small{Laboratoire PRISME, 8, rue L{\'e}onard de Vinci, 45072 Orl{\'e}ans, FRANCE}
\end{center}


\centerline{\bf Abstract}

\noindent
The performances of an original passive control system based on a biomimetic approach are assessed by investigating
the flow over a bluff-body. This control device consists in a couple of flaps made
from the combination of a rigid plastic skeleton coated with a porous fabric mimicking
the shaft and the vane of the bird's feathers, respectively. The sides of a square
cylinder have been fitted with this system so as to enable the flaps to freely rotate
around their leading edge. This feature allows the movable flaps to self-adapt to
the flow conditions. Comparing both the uncontrolled and the controlled flow, a
significant drag reduction (up to 22\%) has been obtained over a broad range of Reynolds
number. The investigation of the mean flow reveals a noticeable modification of the
flow topology at large scale in the vicinity of the controlled cylinder accounting for the
increase of the pressure base in comparison with the natural flow. Meanwhile, the study of 
the relative motion of both flaps points out that their dynamics is sensitive to the
Reynolds number. Furthermore, the comparative study of the flow dynamics at large scale
suggest a lock-in coupling of the flap motion and the vortex shedding.

\vskip 1truecm

\section{Introduction}
\label{sec:intro}

The flow over a bluff body is a situation often encountered in a large number of engineering applications such as aerodynamics of road vehicles or buildings undergoing wind loading. One of main features of bluff body flows relies on the onset of flow separation arising from either curvature or adverse pressure gradient effects. In most cases, flow separation is responsible for the alteration of the aerodynamic performances (e.g. drag increase) or structural vibrations (e.g. galloping), inducing consequently energy consumption excess and/or structural fatigue. Even though a large number of studies have been devoted to flow separation it remains one of the most challenging issue of modern fluid dynamics.

In this context, flow control appears as one of the most attractive way to prevent flow separation or at least attenuate its effects. At first glance, one can distinguish between two main classes of control strategies: passive and active \cite{Choietal08}. By opposition to the active control, the passive control (often referred to as flow management \cite{GadElHak00}) does not require external power supply. Bearman and Owen \cite{BearmanOwen98} achieved up to 50\% drag reduction of rectangular cylinders by introducing spanwise waviness of the front face. These authors pointed out that the vortex shedding was suppressed for an optimal wavelength, while the separated shear layer instabilities were still observed. Owen et al. \cite{Owenetal01} implemented protuberances at the surface of a circular cylinder following a helical pattern. This passive system enabled 25\% drag reduction over one decade
of Reynolds numbers. Recently, Shao and Wei \cite{ShaoWei08} investigated the modification of the flow around a square cylinder at high Reynolds number by means of a control rod. By displacing the control rod, the authors identified the location zones where the control was efficient. Their results indicate that the maximum drag reduction ($\approx 25$\%) was achieved when the control rod was located in the separated shear layers. In that configuration, the vortex shedding was almost annihilated. Even though these examples, chosen among others, suggest that passive control can lead to significant improvements, the efficiency of this control
strategy is often restricted to limited configurations. For instance, Shao and Wei \cite{ShaoWei08} reported noticeable influences of the Reynolds number and the cross-section form of the control rod onto the control efficiency.
 
Inspired from nature observations, a new class of passive control strategy, referred to as self-adaptive passive control, has recently emerged with the pioneer works of research groups in biomimetism (see e.g. \cite{Bechertetal00}). Indeed, nature provides numerous examples of passive or active control mechanisms. One can cite, for instance, the ability of fishes to optimize their hydrodynamic performances by means of riblets or compliance (see \cite{FishLauder06} for a review). Shape reconfiguration is another example of natural flow management. Alben et al. \cite{Albenetal02} and then after Gosselin et al. \cite{Gosselinetal10} reported huge drag reduction of flexible bodies set in a stream. Recently, Favier et al. \cite{Favieretal09} simulated the flow over a circular cylinder fitted with movable cilia at low Reynolds number. The interplay between the coating and the large scale structure in the wake implied up to 15\% drag reduction. The authors observed a lock-in phenomenon of the coating motion at a frequency slightly smaller than the
natural (i.e. without cilia) vortex shedding frequency. Gosselin and de Langre \cite{GosselindeLangre11} reported an experimental investigation of the aerodynamics of a sphere fitted with a poroelastic coating. Their results showed a significant drag reduction due to the reconfiguration of the hairy surface.

Bechert et al. \cite{Bechertetal00} mimicked a bird's feather by implementing a movable flap on the upper surface of an airfoil. During their experiments, this flap was activated by the flow separation arising at high enough angle of attack. Their results revealed that using this self-adaptive passive device the stall regime was delayed. This phenomenon was recently confirmed by Schatz et al. \cite{Schatzetal04} who performed numerical simulations implementing comparable movable flap. In this study, we develop and evaluate the performances of an original self-adaptive passive control device based on the works reported by Bechert et al. \cite{Bechertetal00} and Schatz et al. \cite{Schatzetal04}. Our system consists in a couple of porous flaps designed to mimic the main features of bird's feathers. This control system is fitted on the sides of a square cylinder model which is well documented in literature. The experimental set-up is made in such way that the control flaps can freely rotate around their leading edge. This system is therefore activated by the flow separation arising at the corners of the square cylinder \cite{LynRodi94}. The work reported here is dedicated to the investigation of the control system dynamics and its interplay with the flow. A specific attention is given to the aerodynamic performances of the controlled square cylinder.

The paper is organized as follows. The experimental set-up and the measurements techniques are detailed in Sec. \ref{sec:expe}. The main features of the uncontrolled flow are described in Sec. \ref{sec:natural} and will provide a reference case. Then, the design of the original control
system developed in this study is introduced in Sec. \ref{sec:passive}. A particular focus is given to the dynamics of the control system with respect to the flow conditions. Finally, the efficiency of this system is assessed in Sec. \ref{sec:control} over a broad range of flow conditions.

\section{Experimental set-up}
\label{sec:expe}

\subsection{The experimental facility}

\begin{figure}[htbp]
\centering
\includegraphics[width=0.8\textwidth]{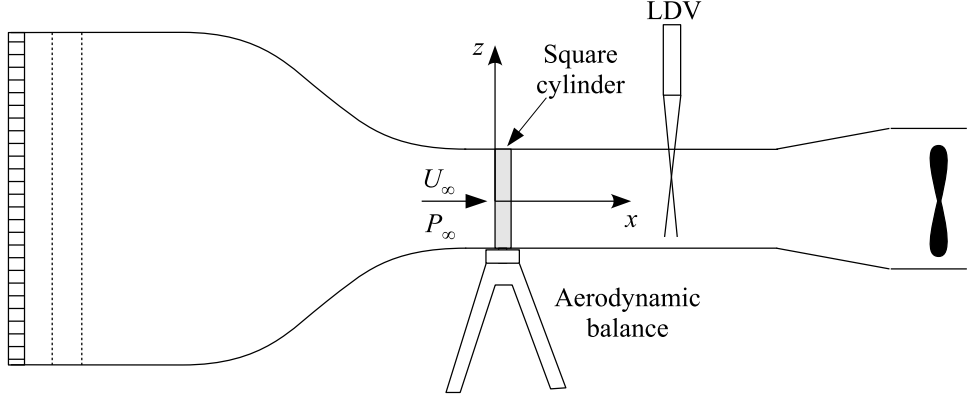}
\caption{Schematic diagram of the experimental set-up.}
\label{fig:Tunnel}
\end{figure}

\begin{figure}[htbp]
\centering
\includegraphics[width=0.70\textwidth]{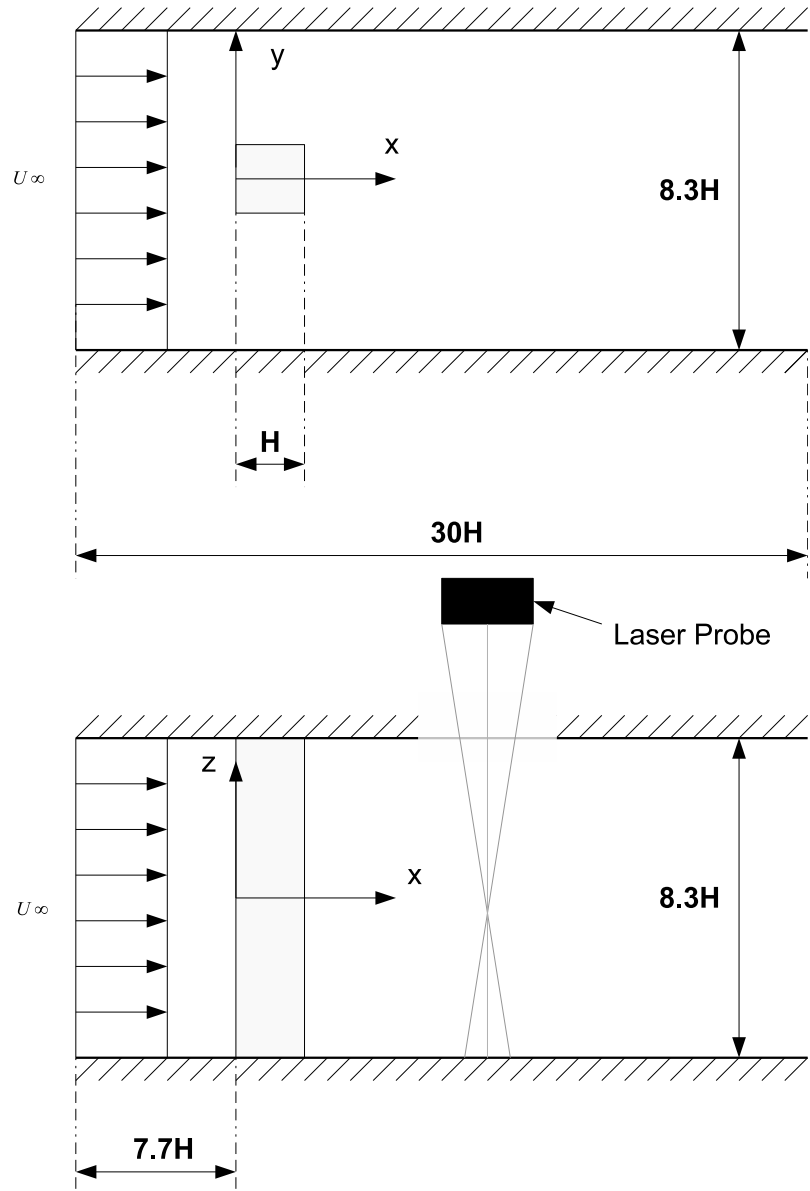}
\caption{Schematic diagram of the working section and the coordinate system.}
\label{fig:WTS2}
\end{figure}

Experiments were conducted in a subsonic open-loop wind tunnel (see Fig. \ref{fig:Tunnel}). The working section is 2m long with a square cross-section of 50cm in width. The combination of a settling chamber equipped with honeycomb and screens followed up with a (16:1) contraction located in front of the working section ensured a very low residual turbulence level ($<0.4$\%).
The free-stream velocity $U_\infty$ is inferred from the pressure drop measured in the contraction. The maximum reachable velocity is 60m/s with the wind tunnel free of obstacle. As shown in Fig. \ref{fig:WTS2}, the roof of the working section is equipped with a movable glass window enabling velocity measurement via an optical technique (see below for more details).

Two aluminum cylinders with a square section were used in the framework of this study. One of these models
was used as a reference, while the other was fitted with the control system introduced in the next section.
The width $H$ of the cylinders was equal to 60mm and their front face were located around 8$H$ downstream
from the working section inlet. The cylinders almost spanned the channel resulting in a blockage ratio
of 12\% and an aspect ratio of 8.3. In this study, the Reynolds number $Re = U_\infty H / \nu$ (with
$\nu$ the kinematic viscosity) was varied over the range $2\cdot10^4$ to $8\cdot10^4$. The origin of the
coordinate system used in the following coincides with the center of the front face of the cylinder as displayed
in Figs. \ref{fig:Tunnel} and \ref{fig:WTS2}. The axis of each cylinder was aligned with gravity (i.e.
$z$-direction here).

\subsection{The measurement techniques}

The velocity field in the vicinity of the square cylinder is investigated by means of two-components (514.5nm, 488nm) Laser Doppler Velocimetry (LDV) system with a 6W Argon-Ion laser (Spectra-Physics, Stabilite 2017) as light source. The flow seeding was ensured by saturating the wind tunnel room with olive oil droplets ($\sim 1 \mu$m in
diameter) generated by a particle seeding apparatus. The LDV probe was mounted on a 3D
traversing system controlled by a computer and data were collected in the backward mode
and processed with a BSA processor (Dantec) set in a non-coincident single measurement
per burst mode. Velocity measurements were performed in the horizontal mid-plane (i.e.
$z=0$) of the working section. The constraints induced by the model geometry and the
set-up configuration prevented us to probe the wall normal velocity closer than
$0.3H$ from the cylinder surface, while the wall tangential velocity was probed up to
$0.05H$ from this surface. The characteristic
length scales of the measurement volume were $80\mu m$ in $x-$ and $y-$directions and
1mm in $z-$direction. For each measurement point, several parameters such as the photomultiplier
intensity and/or the window filtering were adjusted to optimize the mean data rate.
Regarding the investigated region of the flow, the sampling rate was ranging from 2kHz to
20kHz resulting in time-series lasting between 50s (close to the cylinder) and 5s (in the free
stream), respectively. These values were high enough to resolve the large scale flow dynamics.
The LDA technique is known to produce non-equidistant time samples avoiding consequently
a direct spectral analysis (Fourier transform). To overcome this drawback, a linear interpolation
method was used to recover uniformly sampled data in order to investigate the large scale
dynamics of the flow since this procedure only alters the high frequency range. 

The global aerodynamics efforts acting on the square cylinder were obtained using a six
internal components balance which was calibrated separately with masses covering the entire
range of force measured in this study. The force signal was collected
by means of a 12-bits data acquisition card with a sampling rate set at 10Hz. For each
measurement, the integration time was equal to 60s. The time averaged force was then calculated
by subtracting the residual force recorded with flow at rest over 30s before and after
each measurement point. The uncertainties were estimated to lie within the range 2\% (highest $Re$)
to 10\% (lowest $Re$).

Besides the force measurement, the pressure distribution around the square cylinder was
investigated by means of 46 taps distributed in the mid-span of the model. Pressure
taps were connected to a multiplexed pressure transducer (Scanivalve, CTRL2/S2-S6) through
1m long tubing of 1.6mm inner diameter. The pressure reference, $P_\infty$, measured at the 
inlet of the working section was also connected to the pressure transducer such that the
collected signal was the differential pressure $\Delta P = P - P_\infty$. To improve the signal
to noise ratio, the pressure signal was amplified and then low-pass filtered before being stored
via a 16-bits data acquisition card. The sampling rate and the integration time were set at 1kHz
and 30s, respectively.

\section{The natural or uncontrolled flow}
\label{sec:natural}

This section is dedicated to the investigation of the flow around the square cylinder without
the control system. In the following, this configuration is referred to as either the natural
or the uncontrolled flow.

\begin{figure}[htbp]
\centering
\includegraphics[width=0.7\textwidth]{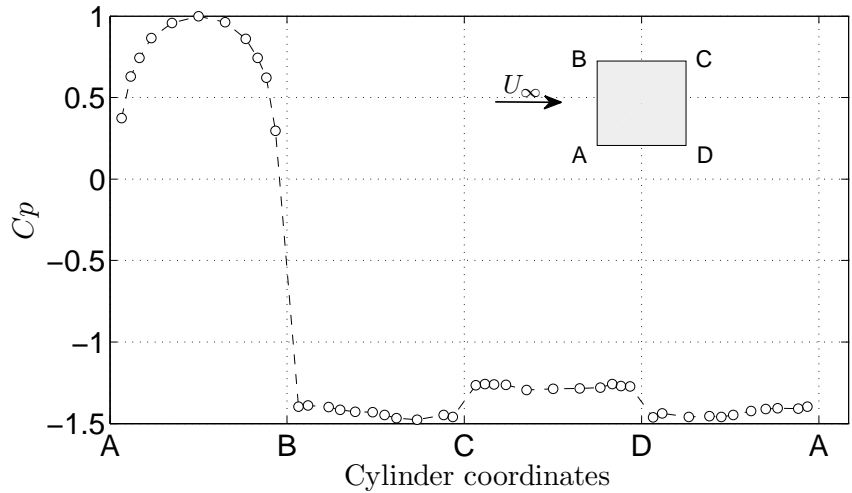}
\caption{Pressure coefficient distribution around the uncontrolled square cylinder
($Re = 2 \cdot 10^4$). The insert illustrates the notation used to describe the spatial
location of the pressure taps around the cylinder.}
\label{fig:CpWoF}
\end{figure}

It is well known that the forces acting on bluff bodies are mainly dominated by pressure forces
\cite{Roshko55} unlike streamlined bodies where viscous forces are not negligible. Therefore, in
this study, the pressure coefficient $Cp = \Delta P / (\rho U_\infty^2 / 2)$ is used as a
relevant variable allowing for the characterization of the local fluid force exerted on the
square cylinder. The distribution of $Cp$ around the uncontrolled cylinder measured for $Re = 2
\cdot 10^4$ is displayed in Fig. \ref{fig:CpWoF}. For sake of simplicity, the faces of the square
cylinder have been decomposed as the front face (AB), the base (CD) and the sides (BC and DA).
It is worth noticing that the pressure distribution is symmetric with
respect to the $y$ axis meaning that the mean lift is null. Therefore, for convenience, we only
focus on the half cylinder (i.e. $y \geq 0$) in the following.

\begin{figure}[htbp]
\centering
\includegraphics[width=0.9\textwidth]{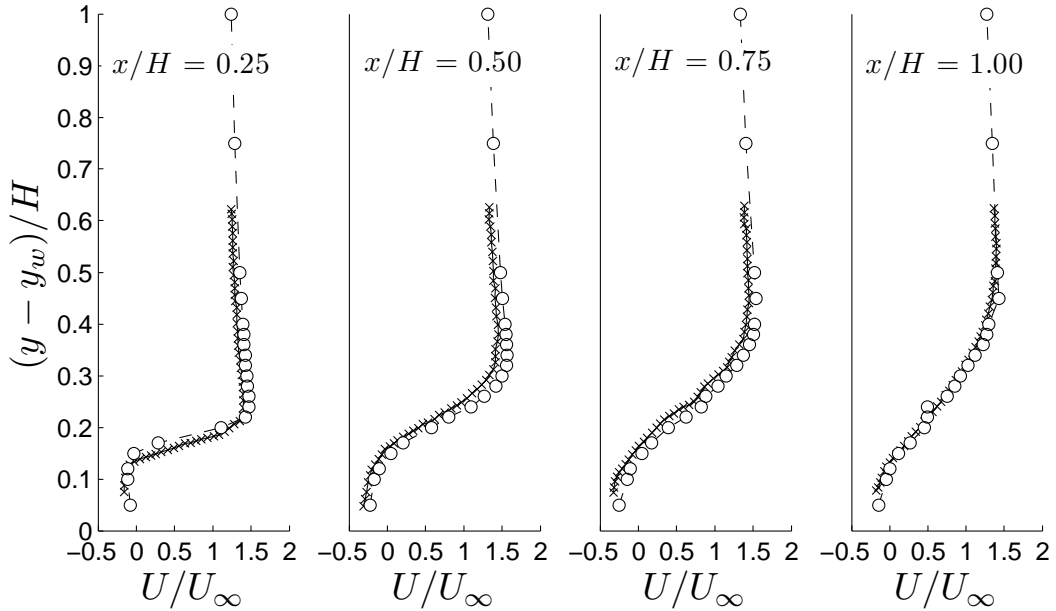}
\caption{Dimensionless streamwise mean velocity profiles $U/U_\infty$ measured on the side BC at
several distances downstream from the corner B ($Re = 2 \cdot 10^4$). The vertical axis represents
the distance from the wall $y-y_w$ normalized by $H$. Our results (open circles) are compared to
those reported by Lyn and Rodi \cite{LynRodi94} at similar $Re$ (crosses).}
\label{fig:UProfileWoF}
\end{figure}

As expected, the front face AB is featured by a strong pressure level where
the maximum value $Cp = 1$ corresponds to the stagnation pressure. The sudden
pressure drop observable at corner B results from the flow separation due to the sharp
curvature of the model. This flow separation induces the formation of a recirculation region on
the side of the square cylinder. This is well supported by the profiles of the normalized streamwise
mean velocity $U / U_\infty$ plotted in Fig. \ref{fig:UProfileWoF} which exhibit a region close to the
side wall (here BC) where the streamwise velocity is negative. Note that for convenience, we introduce
the wall location $y_w$ such that $y - y_w$ stands for the distance from the wall. Our results fairly well agree
with those of Lyn and Rodi \cite{LynRodi94} which have been added in Fig. \ref{fig:UProfileWoF}
for comparison.

\begin{figure}[htbp]
\centering
\includegraphics[width=0.7\textwidth]{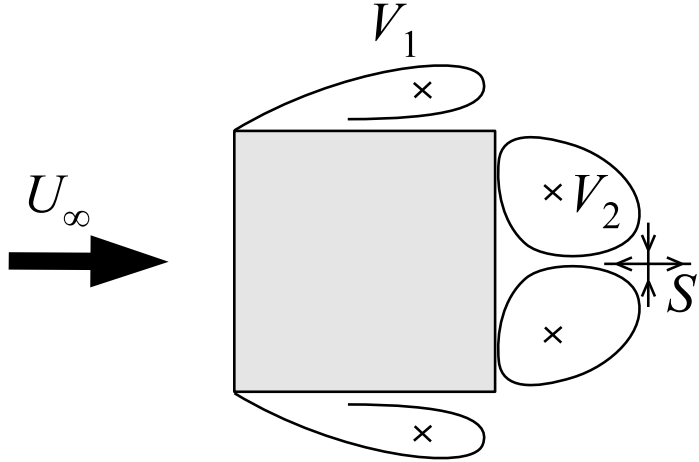}
\caption{Schematic of the flow topology in the vicinity of the uncontrolled square cylinder. The
position $V_1$ stands for the average location of the vortex core of the recirculation regions
on the side. The positions $V_2$ and $S$ denote the average locations of the vortex core and
the stagnation point characterizing the recirculation region on the base, respectively.}
\label{fig:WoF}
\end{figure}

The typical size of the recirculation region may be characterized by the location of the velocity
bulge observable for the three first streamwise distances in Fig. \ref{fig:UProfileWoF}. One can see
that the recirculation region grows with increasing distance $x/H$ until $x/H = 0.75$ which coincides
with the minimum pressure level (see Fig. \ref{fig:CpWoF}). This specific position is representative
of the location of the vortex core ($V_1$, hereafter) featuring the recirculation region on the cylinder
side as illustrated in Fig. \ref{fig:WoF}.

The low pressure plateau on the base (CD) of the model is due to the presence of another recirculation
region. It is important to note that the drag force is induced by the asymmetry of the pressure distribution
between the front and the base faces. The near wake of the uncontrolled cylinder can be characterized
by the points $V_2$ and $S$ introduced in Fig. \ref{fig:WoF}. The former represents the vortex core
position of the flow recirculation on the base, while the latter stands for the location of the stagnation
point delineating the recirculation region in the wake.

\begin{figure}[htbp]
\centering
\includegraphics[width=0.7\textwidth]{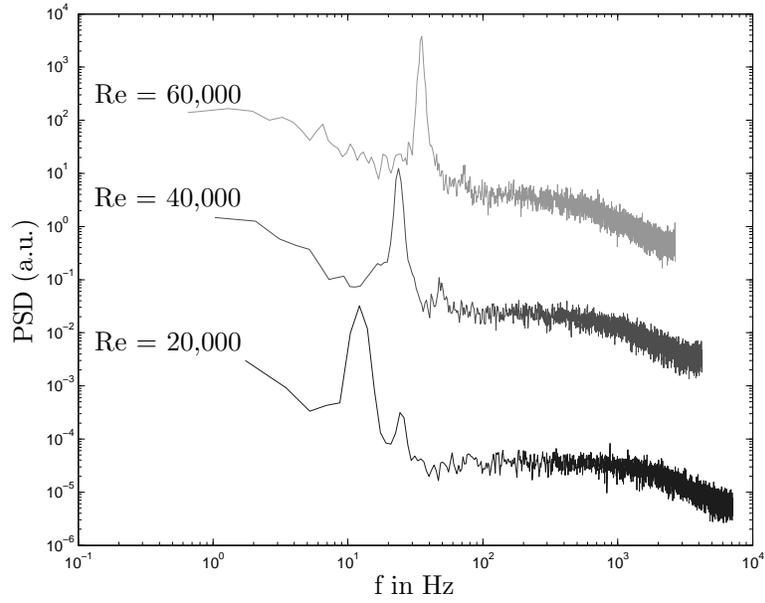}
\caption{Power spectra of velocity as functions of the frequency $f$ computed in the near
wake ($x/H = 2$, $y/H = 2.4$) of the uncontrolled square cylinder for several $Re$. For clarity,
the curves have been shifted along the vertical axis.}
\label{fig:DSPWoF}
\end{figure}

In addition to the mean flow topology, the dominant dynamics of the flow around the uncontrolled
square cylinder has also been studied. The natural vortex shedding frequency of the flow, $f_n$,
was determined using a spectral analysis of the velocity signal as illustrated in Fig.
\ref{fig:DSPWoF} for several $Re$. The related natural Strouhal number $St_n$ ($\equiv
f_n H/U_\infty$) remains almost constant ($St_n = 0.143 \pm 0.002$) over the range of $Re$ used
here. This value falls into the range reported in literature (see e.g. \cite{Norberg93}).

\section{The passive control system}
\label{sec:passive}

\subsection{The design}

\begin{figure}[htbp]
\centering
\subfigure[]
{\includegraphics[width=0.47\textwidth]{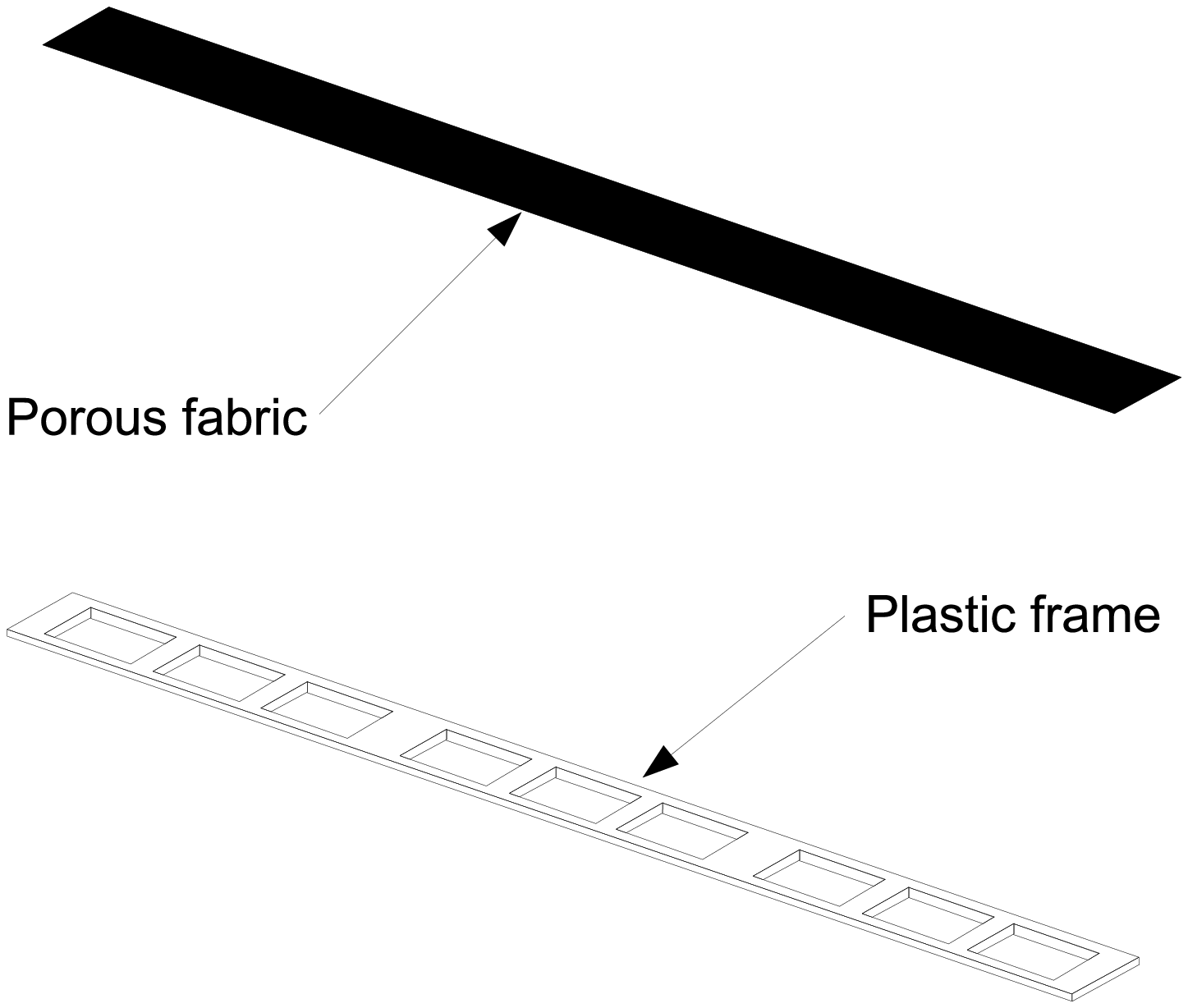}
\label{fig:Device}}
\vspace{0.2cm}
\subfigure[]
{\includegraphics[width=0.3\textwidth]{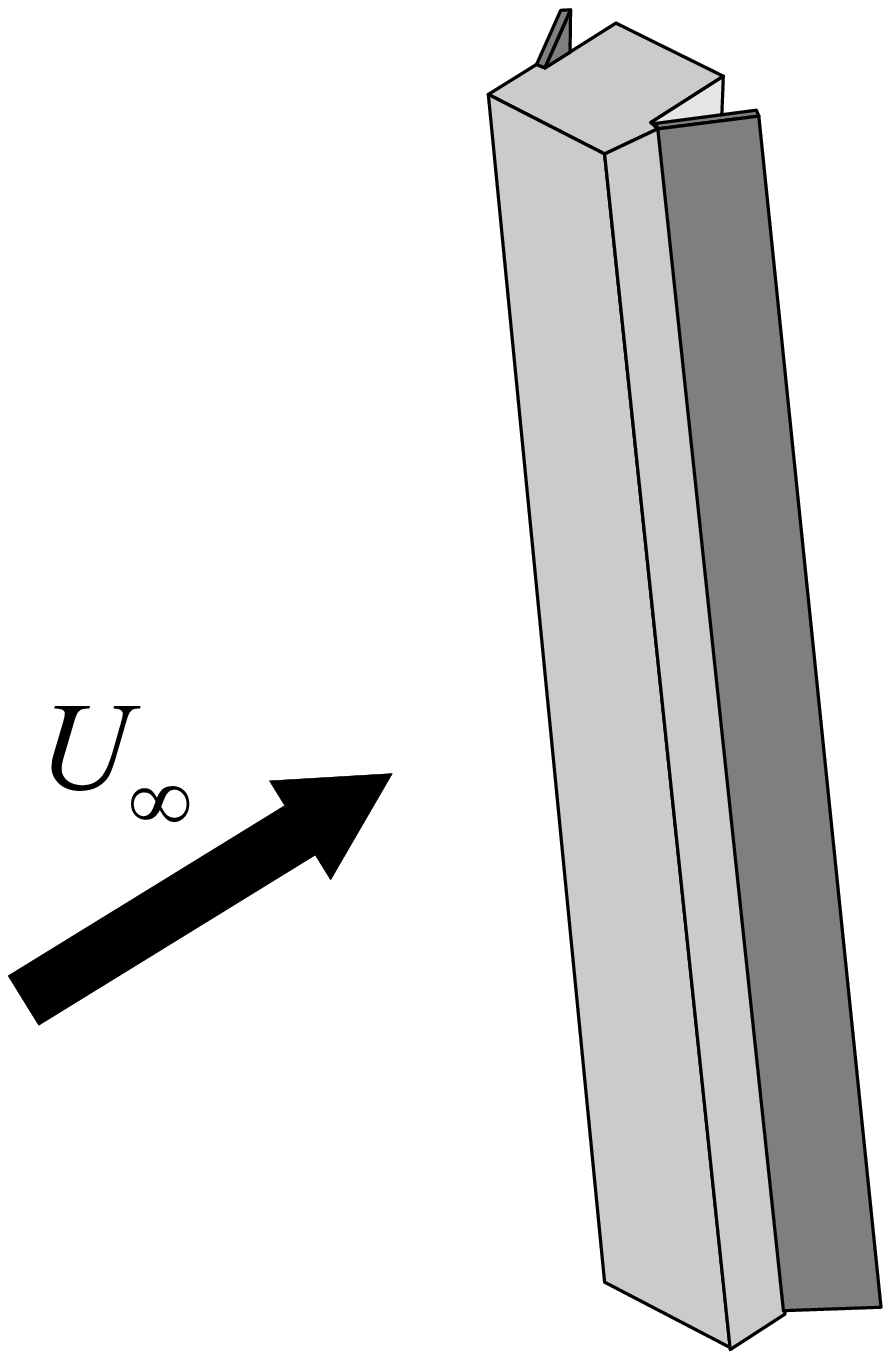}
\label{fig:CylDevice}}
\caption{(a) Exploded view of the self-adaptive control system made from the combination
of a rigid frame (ABS plastic) and a porous fabric (silk). (b) Schematic of the controlled
square cylinder.}
\end{figure}

The control system investigated in this study which consists in a couple of flaps
has been designed to mimic some features of bird's feathers. For that purpose, the
flaps are built from the combination of a rigid frame and a porous fabric simulating
the shaft and the vane, respectively. A schematic of the control flap is given in Fig.
\ref{fig:Device}.

Besides the reference square cylinder, a similar model has been fitted with this original
control system. Each flap is set on one side of the controlled square cylinder at a distance
$x_f$ downstream from the corners A or B. The flaps are fixed to the cylinder by means of
tape such that they can freely rotate around their leading edge. An illustration of
the experimental set-up of the control system is given in Fig. \ref{fig:CylDevice}. In
this study, the chord $c$ and the thickness $e$ of the flaps are equal to 35mm and 2mm,
respectively. Their wingspan $L$ is slightly smaller than the square cylinder span
in order to prevent friction on the wind tunnel walls. Note that all experiments have been
performed with the square cylinder axis aligned with the gravity and that the sides of the
controlled square cylinder have been machined such that, when the fluid is at rest, the flaps
fit into the walls.

According to the results reported in Sec. \ref{sec:natural}, when the fluid passes around
the model, the side walls of the cylinder are subjected to a low pressure level inducing
a suction effect. For high enough $Re$, the suction is sufficiently intense to activate
the movable flaps which depart from the cylinder surface and then flutter around a mean position.
As no external power supply is required to activate our control system, the latter is referred
to as a self-adaptive passive control device. The rigid frame which is made from solid ABS
(rapid prototyping) has been designed to stiffen the flap in order to prevent waviness during
the flap motion. Meanwhile, in order to lighten the flap, the ratio of the area covered by
its solid structure to the reference surface $c L$ has been restricted to about 50\% (see
Fig. \ref{fig:Device}). The porous fabric which is used to mimic the vane of the bird's
feather is made from a commercial silk covering the entire flap area $c L$. To avoid the
propagation of spanwise waves, the fabric is glued onto the solid frame.

Note that even though three different positions of the fixation $x_f$ have been tested,
we only report the results obtained for $x_f = $25mm which has been found to be the
most efficient. It is worth noticing that at this specific position, the center of the
flaps roughly coincides with the location of the vortex core $V_1$ introduced in Fig.
\ref{fig:WoF}.

\subsection{The flap dynamics}

In order to understand the coupling between the flow and the control system, the dynamics
of the movable flaps has been investigated. For that purpose, the time evolution of the
position of the flaps has been evaluated on images recorded by means of a camera with a
sampling rate of 300 frames per second. This sampling rate is at least 8 times larger
than the highest vortex shedding frequency $f_n$ observed for the natural flow (see
Sec. \ref{sec:natural}).

\begin{figure}[htbp]
\centering
\includegraphics[width=0.9\textwidth]{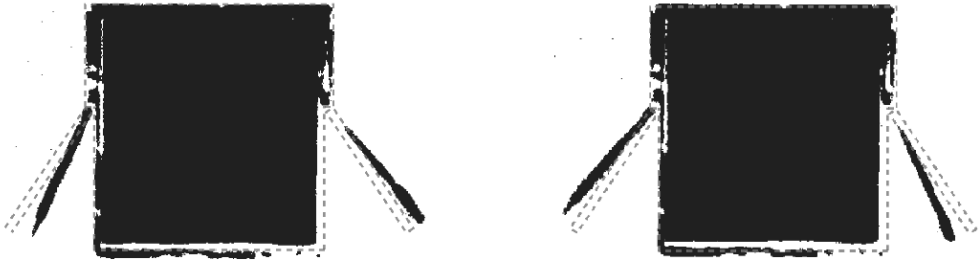}
\caption{Typical snapshots of the movable flaps ($Re = 6 \cdot 10^4$). The dashed
contour symbolizes the average position of the flaps.}
\label{fig:Images}
\end{figure}

The motion of both flaps is illustrated in Fig. \ref{fig:Images} which
shows two typical positions reached by the movable flaps for $Re = 6 \cdot 10^4$. For
comparison, the average position of the flaps has been drawn (dashed contour). One
can see that significant departures from the average position are achieved during the
experiments. Note that for helping the comparison, a threshold procedure has been designed
to convert the original images into binary images.

\begin{figure}[htbp]
\centering
\subfigure[]
{\includegraphics[width=0.45\textwidth]{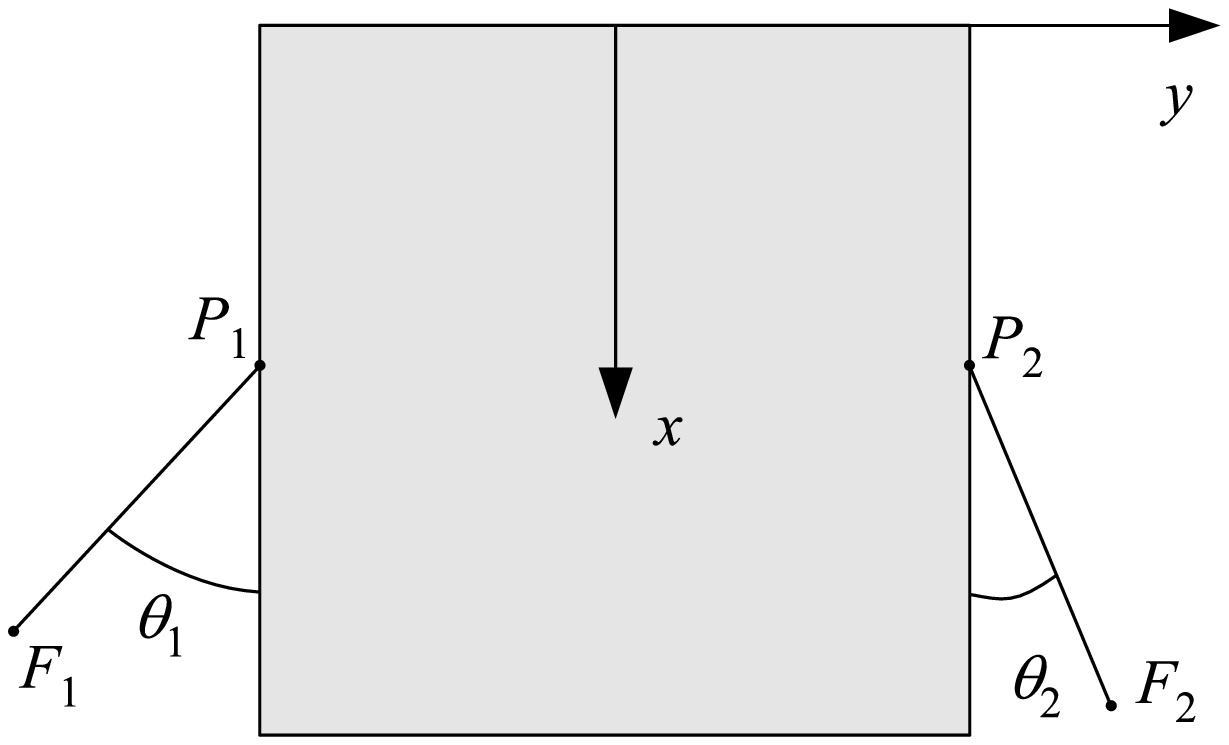}
\label{fig:Angle}}
\hspace{0.2cm}
\subfigure[]
{\includegraphics[width=0.45\textwidth]{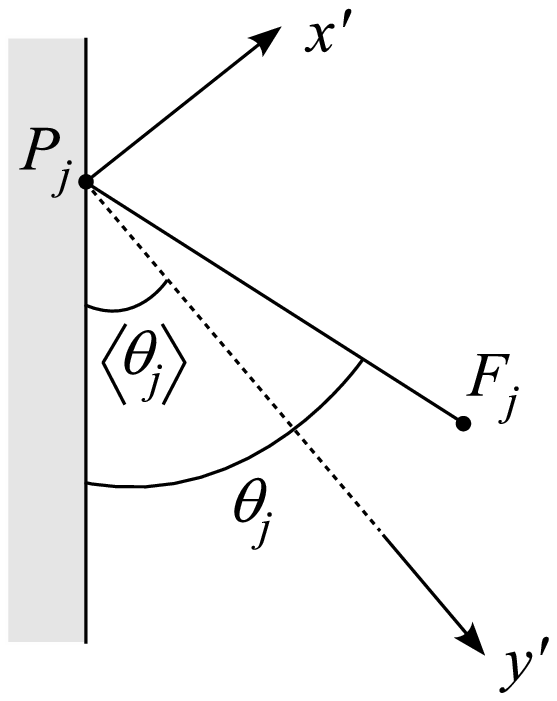}
\label{fig:AngleZoom}}
\caption{(a) Schematic of the geometrical notation used to described the instantaneous
position of the flaps. (b) Local coordinate system attached to the average position of
a flap.}
\end{figure}

The dynamics of the movable flaps is featured by the time evolution of the angle
$\theta_j(t)$ where the subscript $j$ equals either 1 or 2 depending on which flap
is considered. A schematic of the geometrical representation is given in Fig.
\ref{fig:Angle}. Furthermore, we introduce the local coordinate system $\left(x',
y'\right)$ related to the average position of the flaps, i.e. $\left<\theta_j\right>$
(where $\left<\mbox{ }\right>$ stands for the time average), as shown in Fig.
\ref{fig:AngleZoom}. In these diagrams, the points $P_j$ represent the pivot on which
the flaps rotate, while the points $F_j$ denote the instantaneous position of the tip
of the flaps.

A specific algorithm has been developed to recover the time-series $\theta_j(t)$
by evaluating the positions of the flaps on each snapshot. The uncertainties associated
to this procedure has been estimated to be lower than 2\%. The instantaneous location
of the points $F_j$ in the local coordinate system $\left(x', y'\right)$ can therefore
be defined as follows:

\begin{equation}
F_j(t) \equiv
\left\{
\begin{array}{l}
x_j'(t) = c \sin{\left(\theta_j(t) - \left<\theta_j\right>\right)},\\
y_j'(t) = c \cos{\left(\theta_j(t) - \left<\theta_j\right>\right)}
\end{array}
\right.
\label{eq:Traj}
\end{equation}
 
\begin{figure}[htbp]
\centering
\subfigure[]
{\includegraphics[width=0.4\textwidth]{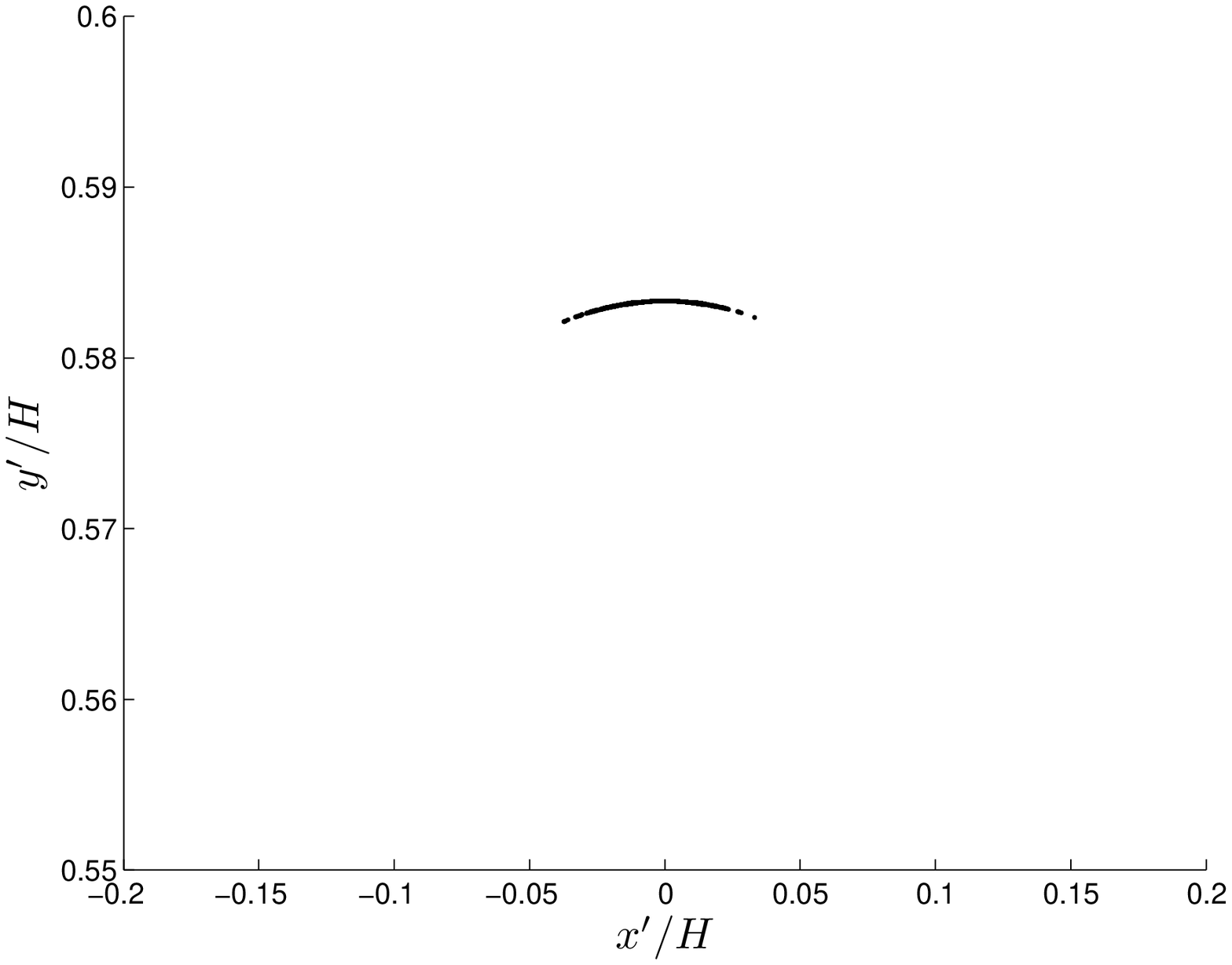}
\label{fig:Traj5ms}}
\hspace{0.2cm}
\subfigure[]
{\includegraphics[width=0.35\textwidth]{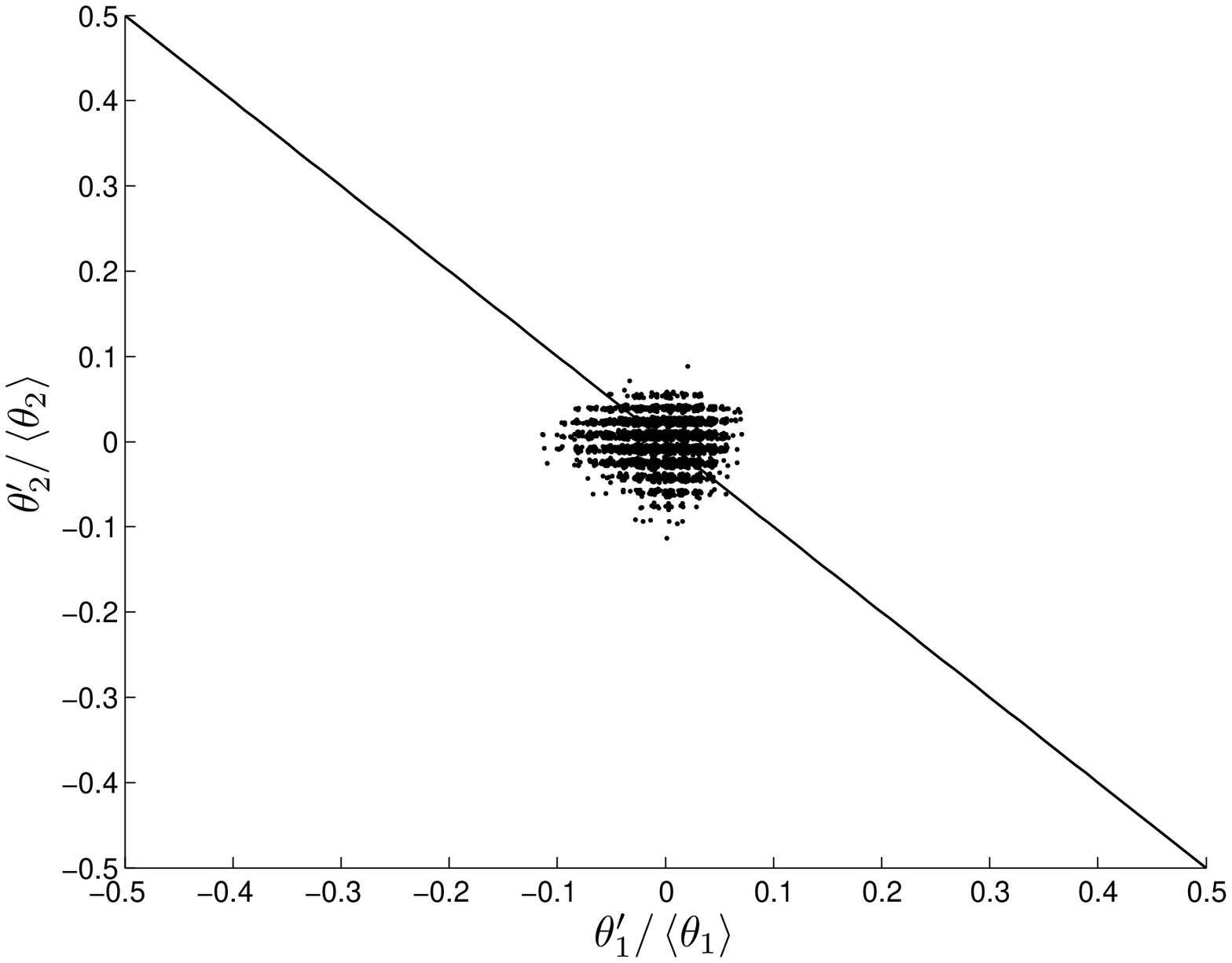}
\label{fig:Scatter5ms}}
\vspace{0.2cm}
\subfigure[]
{\includegraphics[width=0.4\textwidth]{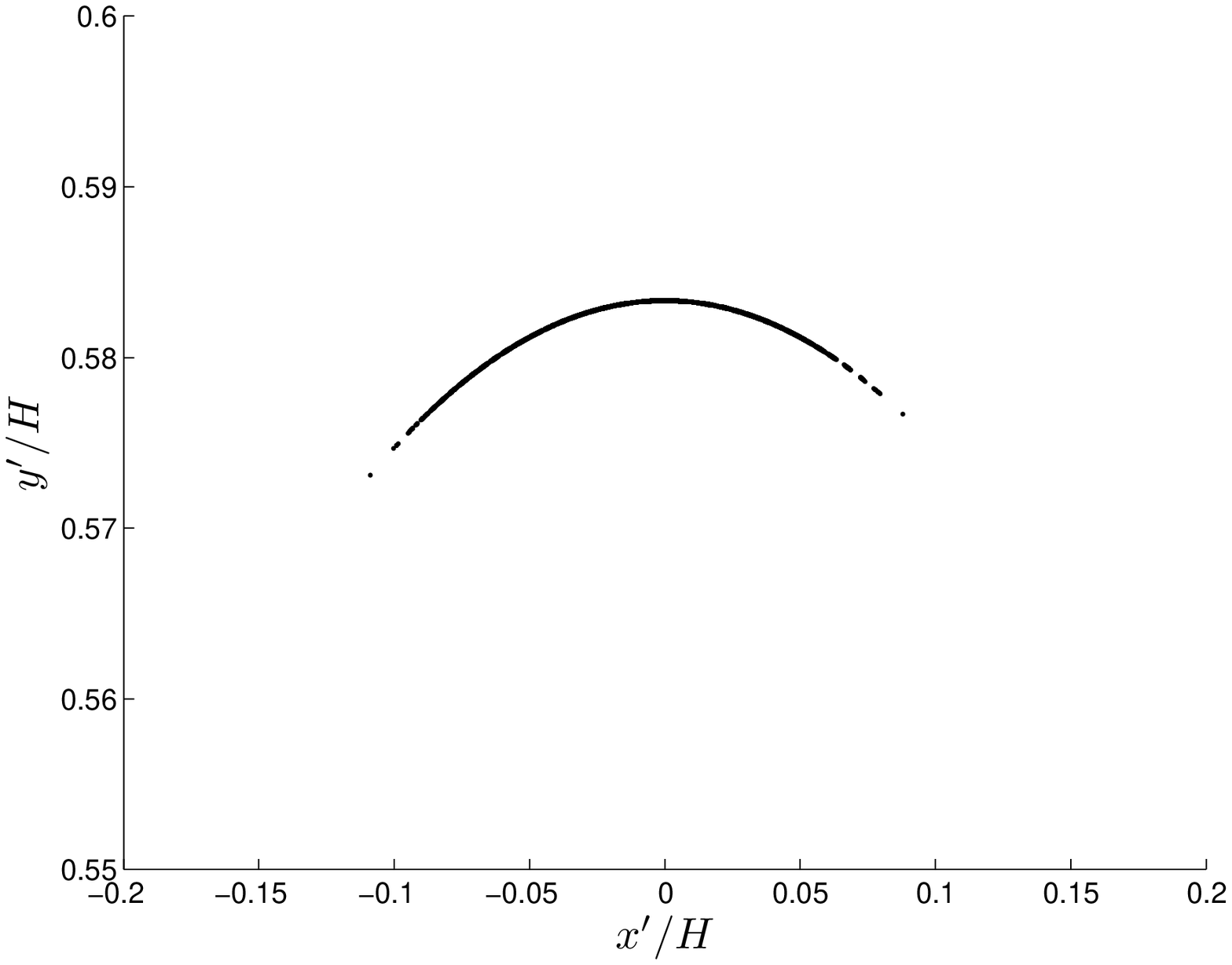}
\label{fig:Traj10ms}}
\hspace{0.2cm}
\subfigure[]
{\includegraphics[width=0.35\textwidth]{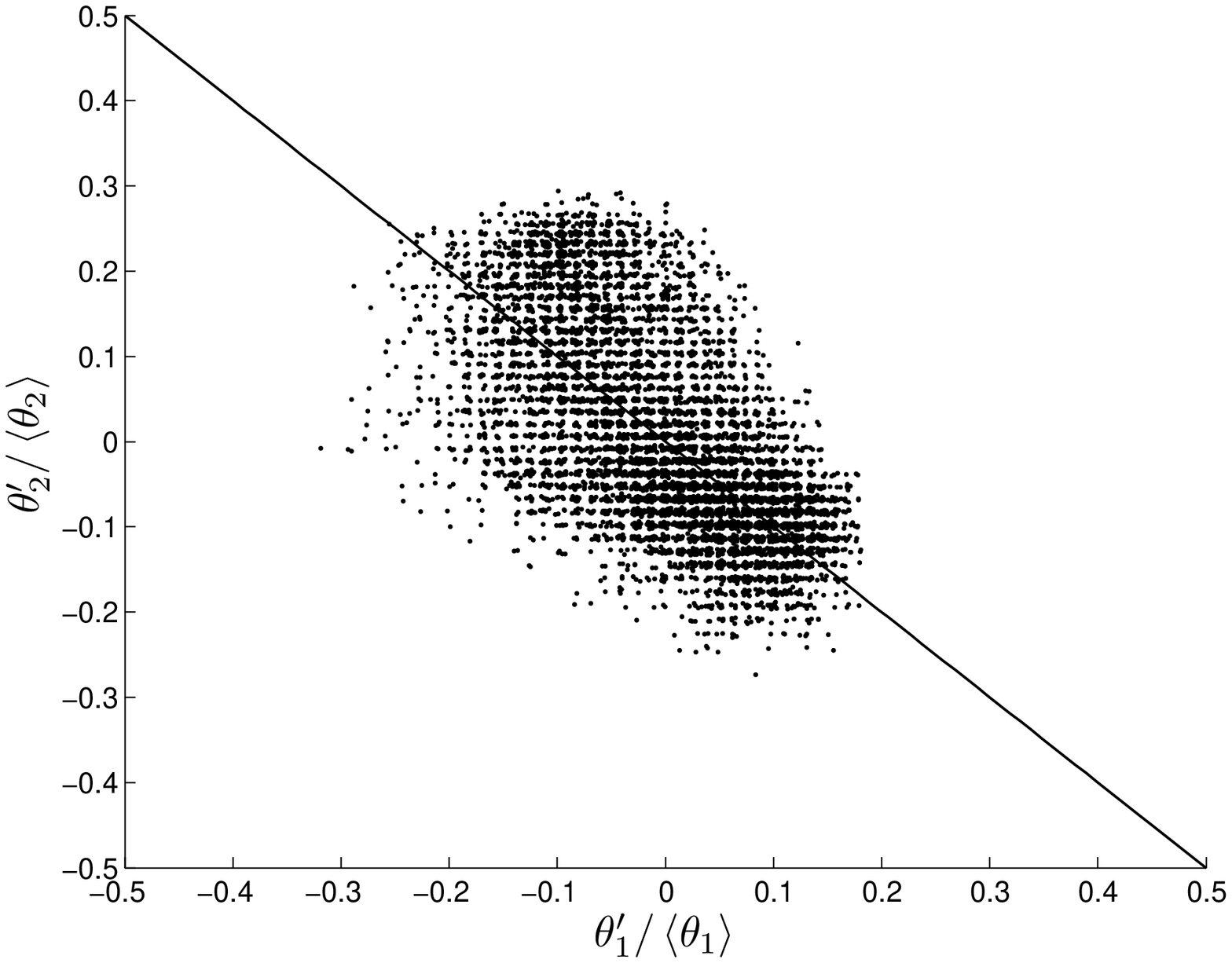}
\label{fig:Scatter10ms}}
\vspace{0.2cm}
\subfigure[]
{\includegraphics[width=0.4\textwidth]{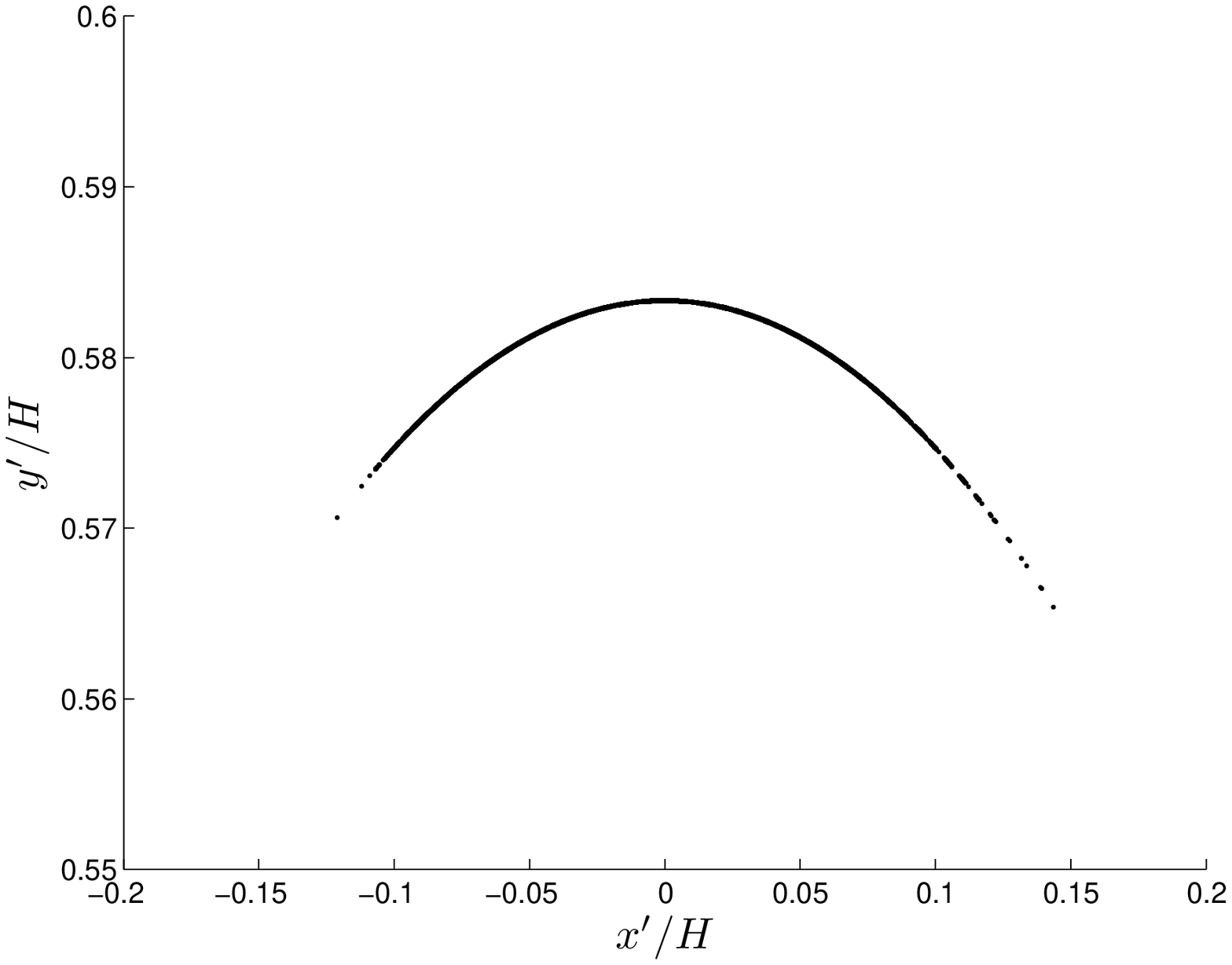}
\label{fig:Traj15ms}}
\hspace{0.2cm}
\subfigure[]
{\includegraphics[width=0.35\textwidth]{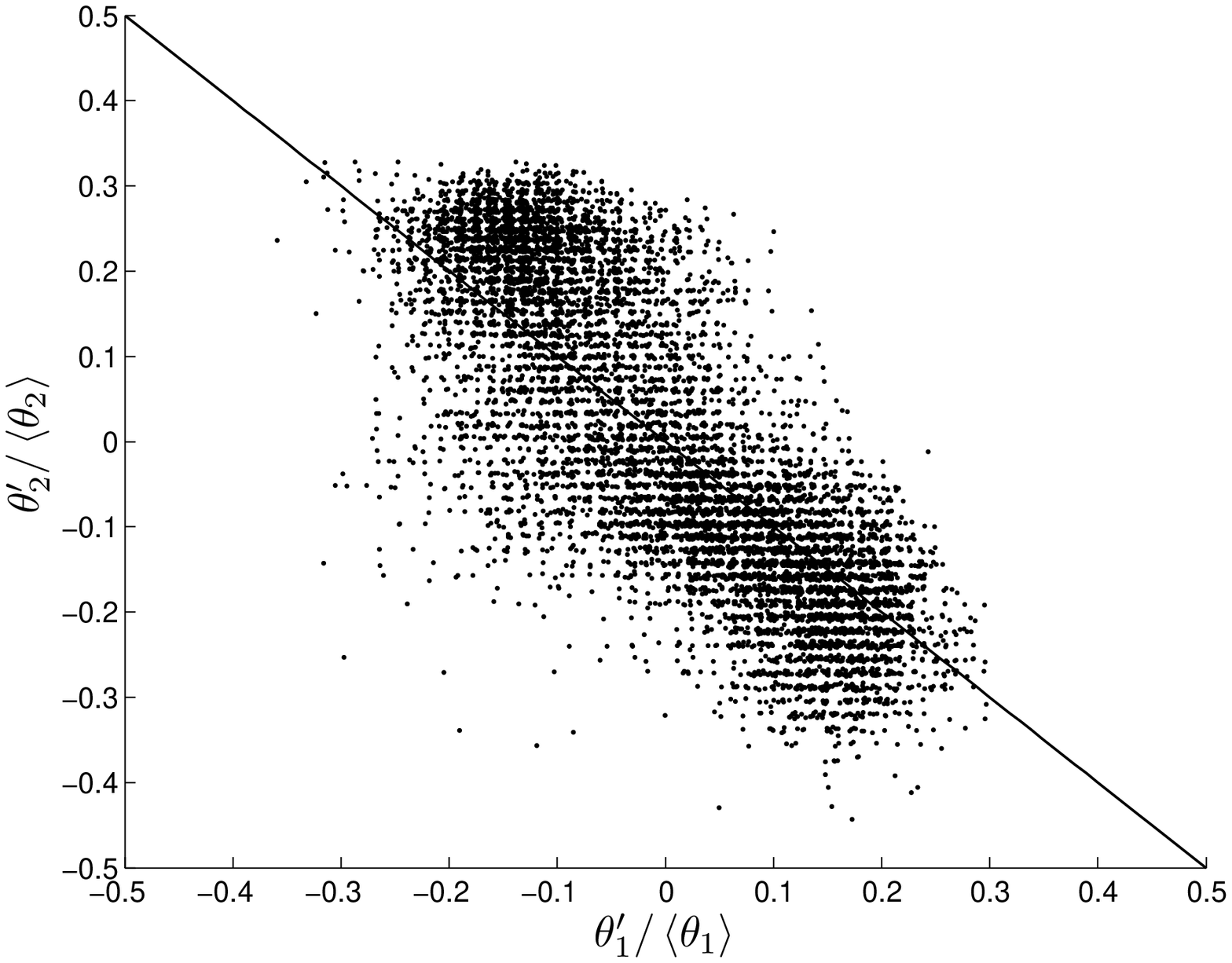}
\label{fig:Scatter15ms}}
\caption{Trajectories of the trailing edge of the movable flaps: (a) $Re = 2 \cdot 10^4$,
(c) $Re = 4 \cdot 10^4$ and (e) $Re = 6 \cdot 10^4$. Scatter plot of the instantaneous angle of
the movable flaps: (b) $Re = 2 \cdot 10^4$, (d) $Re = 4 \cdot 10^4$ and (f) $Re = 6 \cdot 10^4$.
The straight line represents the curve $\theta'_2/\left<\theta_2\right> = -\theta'_1/
\left<\theta_1\right>$.}
\end{figure}

The dimensionless trajectories of the points $F_j$ calculated according to Eq.
(\ref{eq:Traj}) for $Re = 2 \cdot 10^4$,
$Re = 4 \cdot 10^4$ and $Re = 6 \cdot 10^4$ are plotted in Figs. \ref{fig:Traj5ms},
\ref{fig:Traj10ms} and \ref{fig:Traj15ms}, respectively. These results indicate that the
amplitude of the rotation around the equilibrium state, i.e. $\left<x'_j\right>/H = 0$
and $\left<y'_j \right>/H = c/H$, increases with increasing $Re$.

In order to highlight this behavior, we introduce the fluctuating angle $\theta'_j(t)$ such as
$\theta'_j(t) = \theta_j(t) - \left<\theta_j\right>$. The deviation from the equilibrium
state can therefore be quantified by the scatter plots of the normalized fluctuating angles
$\theta'_j(t)/\left<\theta_j\right>$ as displayed in Figs. \ref{fig:Scatter5ms}, \ref{fig:Scatter10ms}
and \ref{fig:Scatter15ms} for $Re = 2 \cdot 10^4$, $Re = 4 \cdot 10^4$ and $Re = 6 \cdot 10^4$,
respectively. For the lowest $Re$, the deviation from the equilibrium angle remains smaller
than 10\%, while it reaches up to 30\% for the highest $Re$. Furthermore, these results point
out a noticeable modification in the shape of the scatter plots with respect to $Re$. The
elongation of the scatter plot observable beyond $Re = 4 \cdot 10^4$ indicates that the motion
of both flaps become dependent to each other. Indeed, the alignment of the scatter plots
along the slope -1 (solid lines in Figs. \ref{fig:Scatter5ms}, \ref{fig:Scatter10ms}
and \ref{fig:Scatter15ms}) means that the rotation of both flaps is in phase.

\begin{figure}[htbp]
\centering
\subfigure[]
{\includegraphics[width=0.45\textwidth]{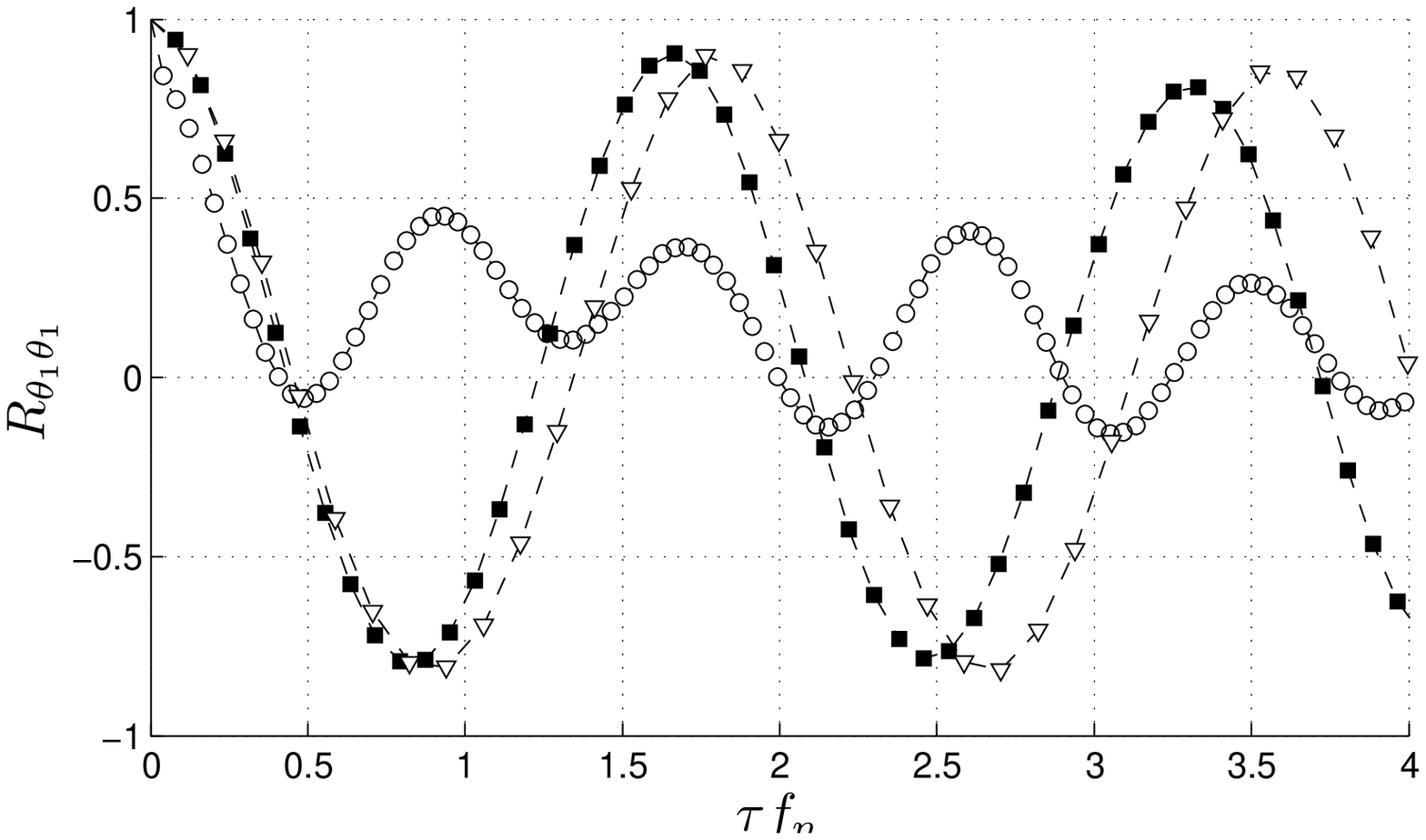}
\label{fig:Corr}}
\hspace{0.2cm}
\subfigure[]
{\includegraphics[width=0.45\textwidth]{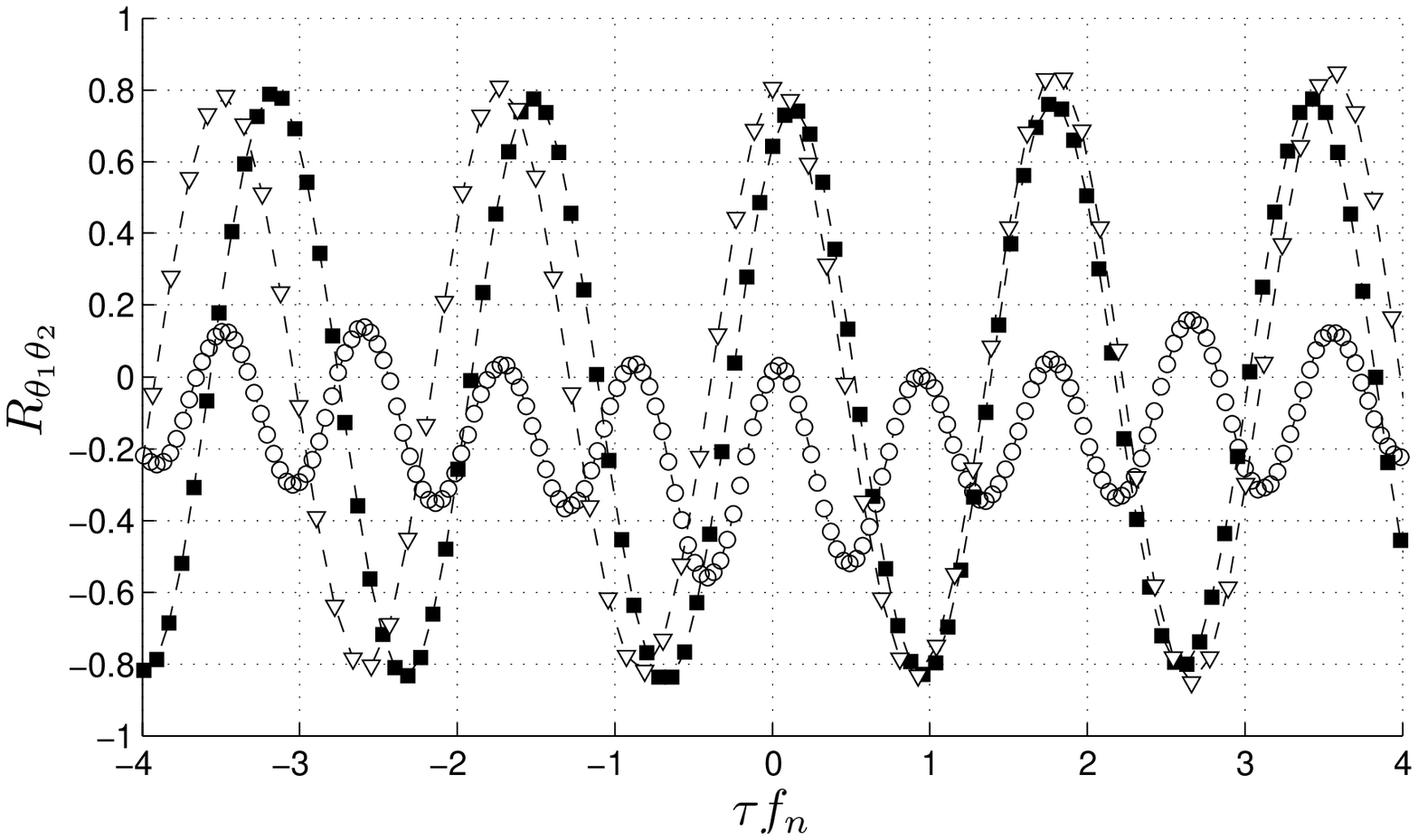}
\label{fig:XCorr}}
\caption{Evolution of (a) $R_{\theta_1\theta_1}$ and (b) $R_{\theta_1\theta_2}$ vs. the dimensionless
time lag $\tau f_n$ for $Re = 2 \cdot 10^4$ (open circles), $Re = 4 \cdot 10^4$ (closed squares) and
$Re = 6 \cdot 10^4$ (open triangles).}
\end{figure}

The dynamics of a single flap can be studied by means of the auto-correlation coefficient
defined as follows:

\begin{equation}
R_{\theta_j\theta_j}\left(\tau\right) =
\frac{\left<\theta'_j(t+\tau) \theta'_j(t)\right>}
{\left<\theta'_j(t)^2\right>},
\end{equation}

where $\tau$ is the time lag. The variation of $R_{\theta_1\theta_1}$ computed for several $Re$ is plotted in Fig.
\ref{fig:Corr}. For sake of clarity, we do not report the variation of $R_{\theta_2\theta_2}$
since no significant difference with $R_{\theta_1\theta_1}$ has been observed. In order to
highlight the differences with the natural (i.e. uncontrolled) flow, the time lag $\tau$ is
normalized by the natural vortex shedding frequency $f_n$ reported previously. The shape of
$R_{\theta_1\theta_1}$ indicates that the rotating motion of the flaps is almost periodic for
each $Re$ used in this study. However, the results reported in Fig. \ref{fig:Corr} show
that the dimensionless rotation frequency of the flaps $f_r / f_n$ is dependent on $Re$. Indeed,
for the lowest $Re$, $f_r$ is slightly higher than the natural vortex shedding frequency ($f_r
\approx 1.2 f_n$), while it is smaller for the two largest $Re$ ($f_r \approx 0.6 f_n$).
We discuss further these results in the next subsection.

In addition to the dynamics of a single flap, the relative motion of both flaps can be quantified
by means of the cross-correlation coefficient:

\begin{equation}
R_{\theta_1\theta_2}\left(\tau\right) =
\frac{\left<\theta'_1(t+\tau) \theta'_2(t)\right>}
{\sqrt{\left<\theta'_1(t)^2\right>}\sqrt{\left<\theta'_2(t)^2\right>}}.
\end{equation}

The variation of $R_{\theta_1\theta_2}$ displayed in Fig. \ref{fig:XCorr} confirms that both flaps
move with the same frequency $f_r$. However, at $\tau f_n = 0$, the cross-correlation
coefficient is almost null for $Re = 2 \cdot 10^4$ meaning therefore that the flaps' motion is
not synchronous. On the contrary, the high value of $R_{\theta_1\theta_2}$ observed at $\tau f_n
=0$ for $Re = 6 \cdot 10^4$, indicate that the flaps rotate in phase. For the intermediate $Re$,
i.e. $Re = 4 \cdot 10^4$, one can see a slight shift in time revealing that the relative flap
rotation is not exactly synchronous.

\subsection{The fluid-structure coupling}

\begin{figure}[htbp]
\centering
\includegraphics[width=0.7\textwidth]{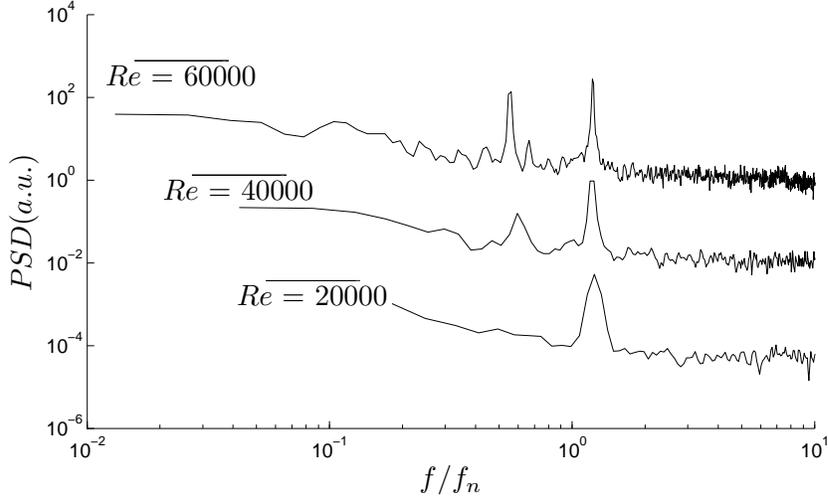}
\caption{Power spectra of velocity as a function of the normalized
frequency $f/f_n$ computed in the near-wake ($x/H = 2$, $y/H = 2.4$) of the controlled square
cylinder for several Reynolds numbers. For clarity, the curves have been shifted along the
vertical axis.}
\label{fig:SpectreArrWF}
\end{figure}

In order to compare the large scale flow dynamics with that of the movable flaps,
we plot in Fig. \ref{fig:SpectreArrWF} the power spectra of velocity measured in
the near wake ($x/H = 2$, $y/H = 2.4$) of the controlled square cylinder. In this plot, the spectral frequency $f$ is normalized by the
natural vortex shedding frequency $f_n$. For the lowest $Re$, i.e. $Re = 2 \cdot
10^4$, the wake is dominated by a single frequency $f_{ci}$ which is comparable to
the rotation frequency $f_r$ of the flaps as shown in the previous subsection. This
frequency features the vortex shedding of the controlled flow. The related
Strouhal number $St_i$ ($\equiv f_{ci} H / U_\infty$) deviates noticeably from the
natural one ($St_i = 0.175 \approx 1.2 St_n$) in agreement with usual observations reported
in fluid/structure interactions (see e.g. \cite{Favieretal09}). For instance, studying
the flow modification around a square cylinder oscillating in a stream, Bearman and
Obajasu \cite{BearmanObasaju82} revealed the occurrence of a lock-in regime between
the vortex shedding and the cylinder oscillation. Such behavior was also reported by
de Langre \cite{deLangre06}  who studied the linear stability of a circular cylinder
subjected to vortex-induced vibrations.
 
For higher $Re$, the wake remains dominated by the vortex shedding characterized by
$St_i = 0.175$ as evidenced in Fig. \ref{fig:SpectreArrWF}. However, another energetic
mode arises at a lower frequency $f_{cg}$. This phenomenon coincides with the synchronization
of the motion of both flaps (i.e. $f_{cg} = f_r$). One can notice that the ratio
$f_{ci}/f_{cg}$ increases with increasing $Re$. Meanwhile, the ratio of the energy contained
in the high frequency mode (i.e. related to $f_{ci}$) to that contained in the low
frequency mode (i.e. related to $f_{cg}$) decreases. This evolution may be interpreted as
a modification of the wake pattern such as reported in vortex-induced vibration (see e.g.
\cite{MorseWilliamson09}) or in active flow control (see e.g. \cite{Pastooretal08}).
However, further experiments and analysis would be required in order to address properly this issue.

\section{The controlled flow}
\label{sec:control}

This subsection is dedicated to the study of the flow subjected to the self-adaptive passive
control. The results obtained for the velocity in the vicinity of the square cylinder and the
pressure distribution around this obstacle are analyzed to provide a first simple mechanism
of the interplay between the control system and the flow. Finally, the efficiency of the control
system is evaluated by means of drag force measurement. 

\subsection{The flow topology modification}

\begin{figure}[htbp]
\centering
\subfigure[]
{\includegraphics[width=0.45\textwidth]{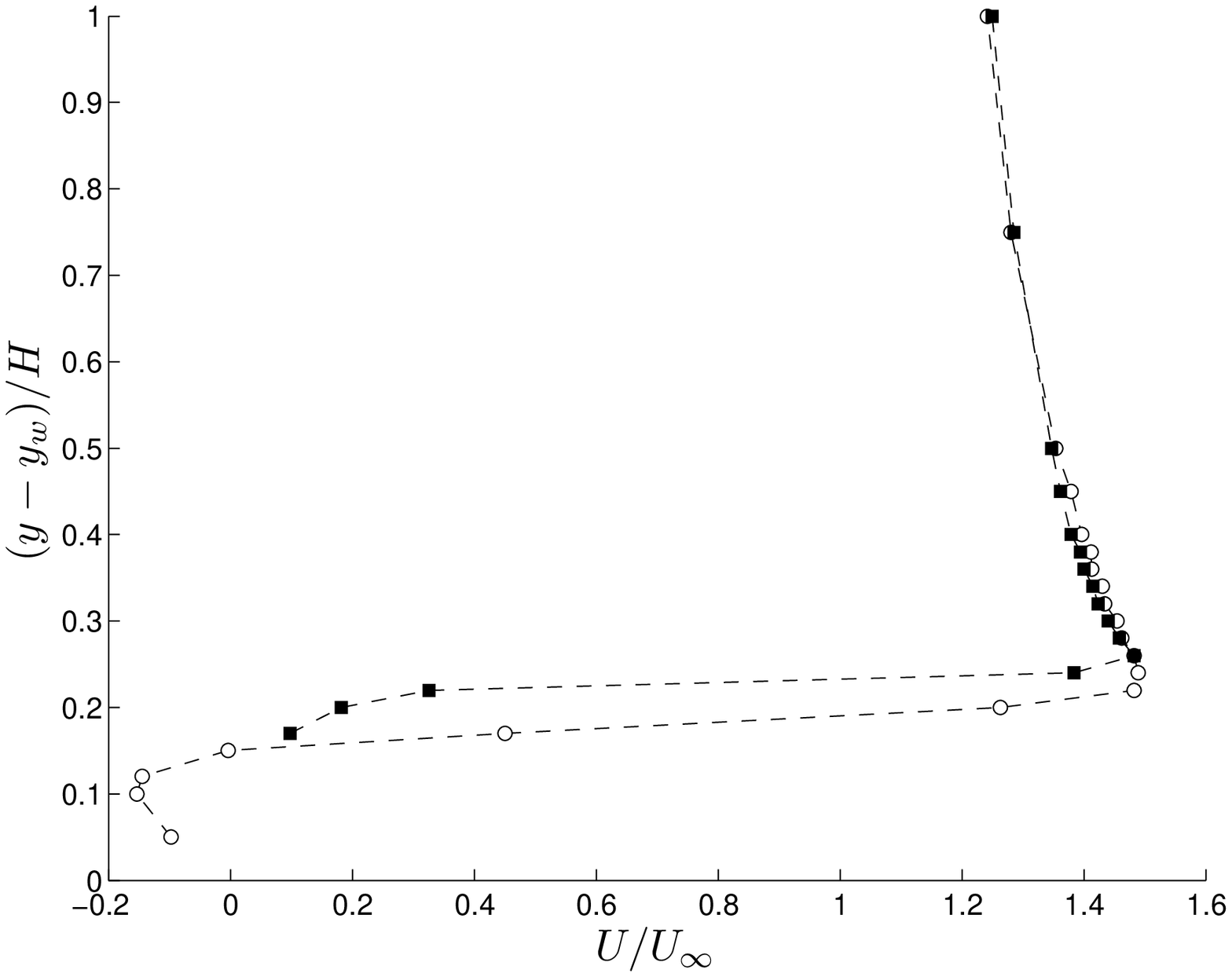}
\label{fig:Side}}
\hspace{0.2cm}
\subfigure[]
{\includegraphics[width=0.45\textwidth]{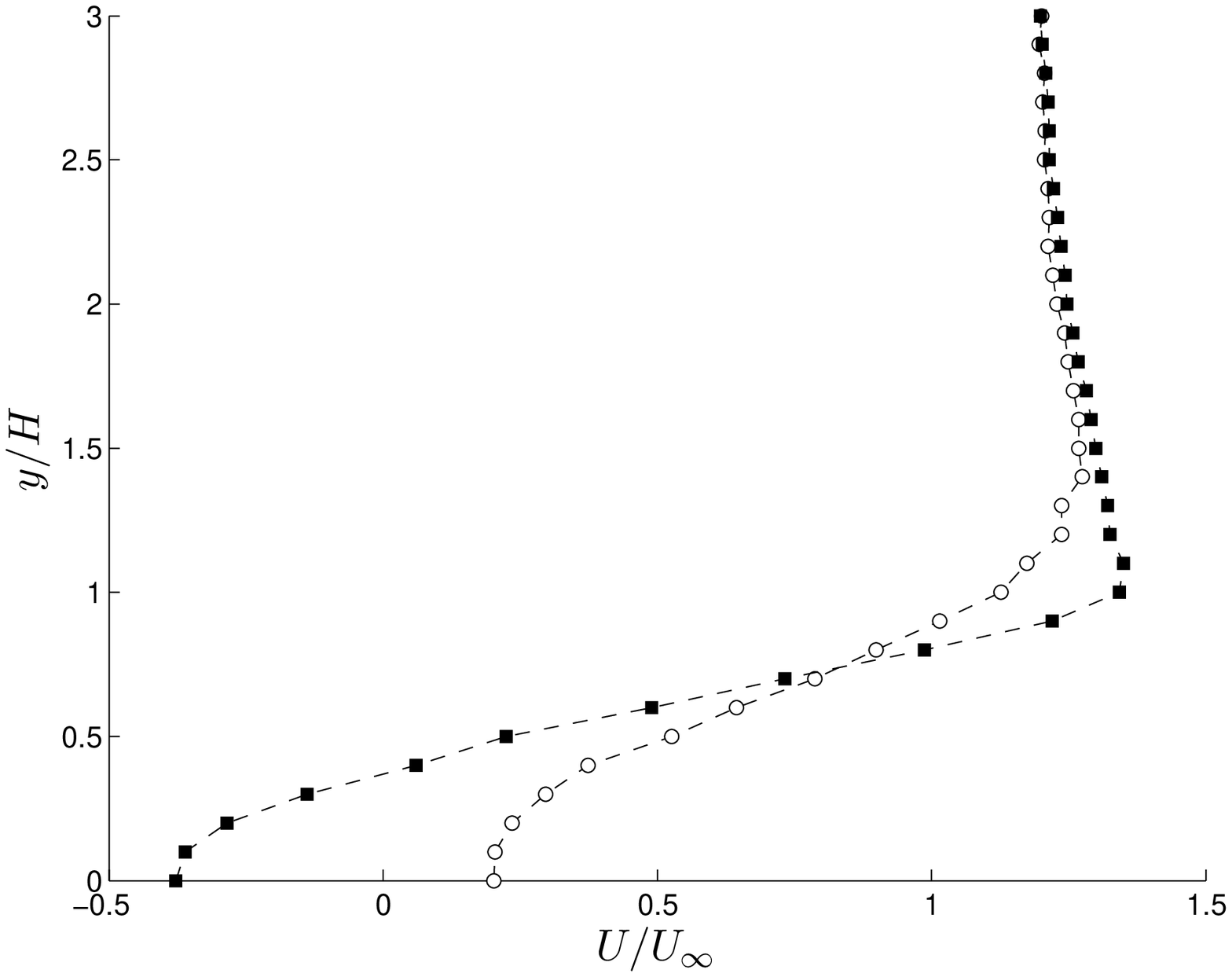}
\label{fig:Wake}}
\caption{Profiles of the dimensionless streamwise mean velocity $U/U_\infty$ measured (a)
on the side at $x/H =0.25$ and (b) in the wake at $x/H = 2$ ($Re = 2 \cdot 10^4$).
The results obtained for both the uncontrolled (open circles) and the controlled (closed
squares) cylinders are compared.}
\end{figure}

Fig. \ref{fig:Side} compares the velocity profiles $U/U_\infty$ normal to the side of both the
uncontrolled and the controlled square cylinders at $x/H = 0.25$ for $Re = 2 \cdot 10^4$. Note
that the crossing of the laser beams and the flap trajectory restricts the investigation of the
velocity field to $(y - y_w) / H \geq 0.17$.

Even though the shape of both curves are similar, one can clearly remark that the recirculation
region is enlarged by the control system. However, beyond $(y-y_w)/H = 0.25$, both velocity profiles
collapse meaning that the outer flow is almost unaffected by the presence of the movable flaps.
This is an important result implying that the blockage of the flow is identical in both the
uncontrolled and the controlled cases.

The modification of the near wake is evidenced in Fig. \ref{fig:Wake} showing the velocity
profile $U/U_\infty$ measured for both the uncontrolled and the controlled square cylinders
at $x/H = 2$ for $Re = 2 \cdot 10^4$. The velocity deficit observed in both cases
is a feature of wakes. However, the spreading of the uncontrolled wake is larger than
that of the controlled one. Moreover, one can see that, at this position, the controlled wake
is featured by a region ($y/H \leq 0.40$) where $U \leq 0$ unlike the uncontrolled wake. This
means that the location of the stagnation point $S$ delineating the recirculation region
in the wake of the cylinder has moved farther downstream in the controlled case.

\begin{figure}[htbp]
\centering
\includegraphics[width=0.7\textwidth]{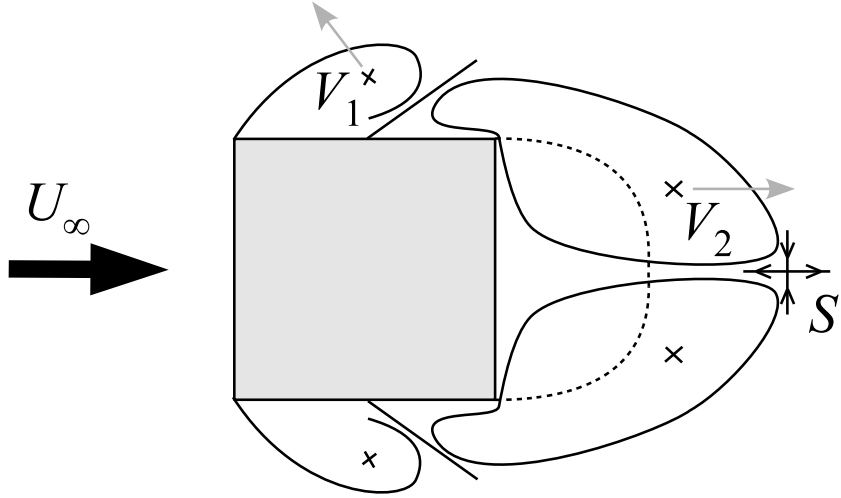}
\caption{Schematic of the flow around the square cylinder fitted with the self-adaptive
movable flaps. The gray arrows symbolize the expected displacement of the vortex cores $V_1$
and $V_2$ relative to their natural position. The dashed line symbolizes the frontier of the
recirculation region in the uncontrolled case.}
\label{fig:WF}
\end{figure}

According to the results displayed in Figs. \ref{fig:Side} and \ref{fig:Wake}, we propose a
simple representation of the mean topology of the controlled flow which is described in Fig.
\ref{fig:WF}. In that representation, two recirculation regions are separated by the movable
flaps. The vortex associated to $V_1$ which is expected to be at the origin of control
activation is pushed upstream by the flap motion. Meanwhile, the pressure level downstream
from the flaps (i.e. right-hand side in Fig. \ref{fig:WF} is expected to increase due to the pressure drop induced by the fluid passing
through the porous fabric. As a consequence, the vortex core $V_2$ of the second recirculation
region, and by the way the stagnation point $S$, are expected to move downstream their positions
in the uncontrolled case. 

\subsection{The drag force reduction}

\begin{figure}[htbp]
\centering
\subfigure[]
{\includegraphics[width=0.6\textwidth]{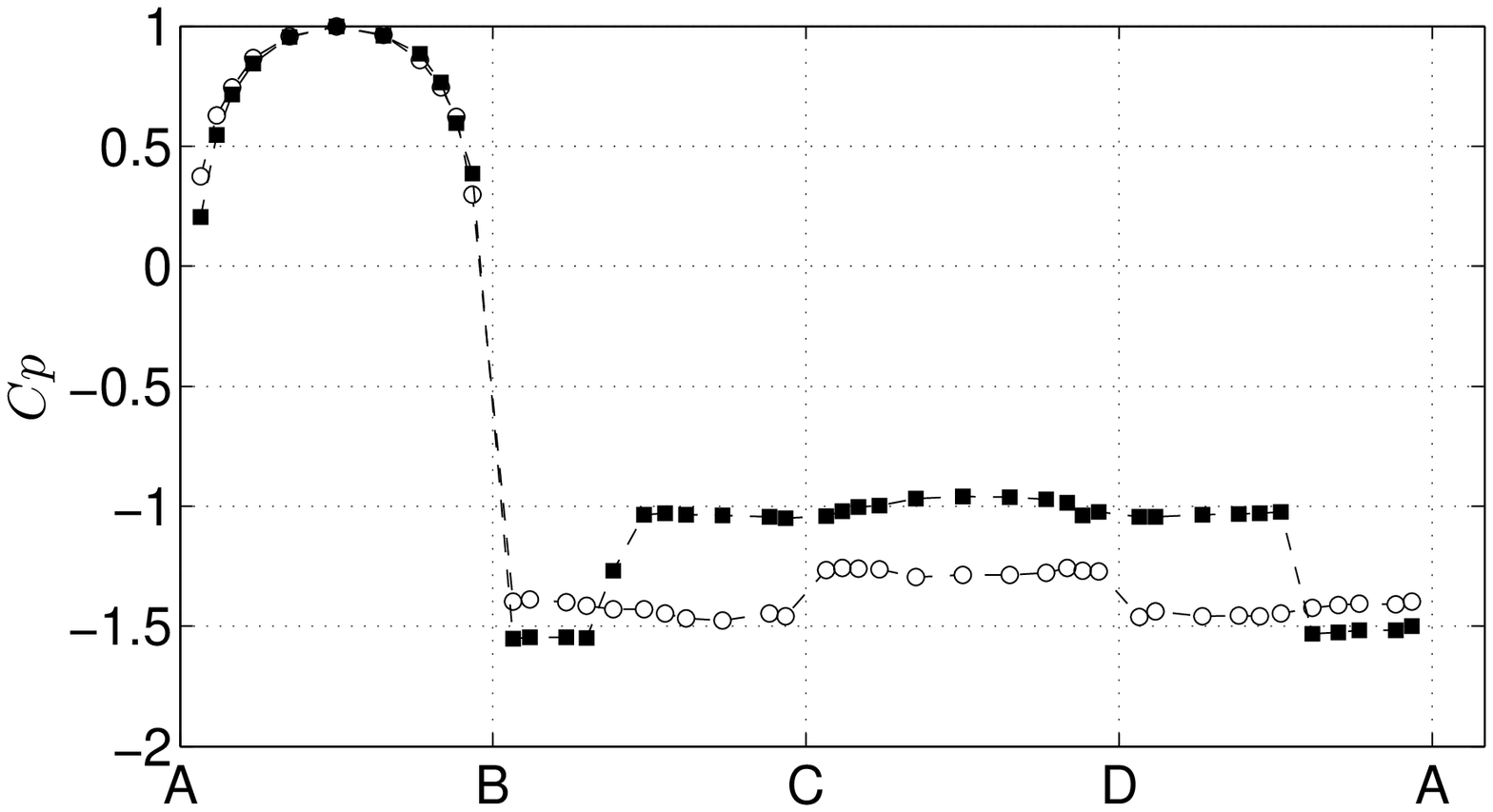}
\label{fig:CpRe20000}}
\hspace{0.2cm}
\subfigure[]
{\includegraphics[width=0.6\textwidth]{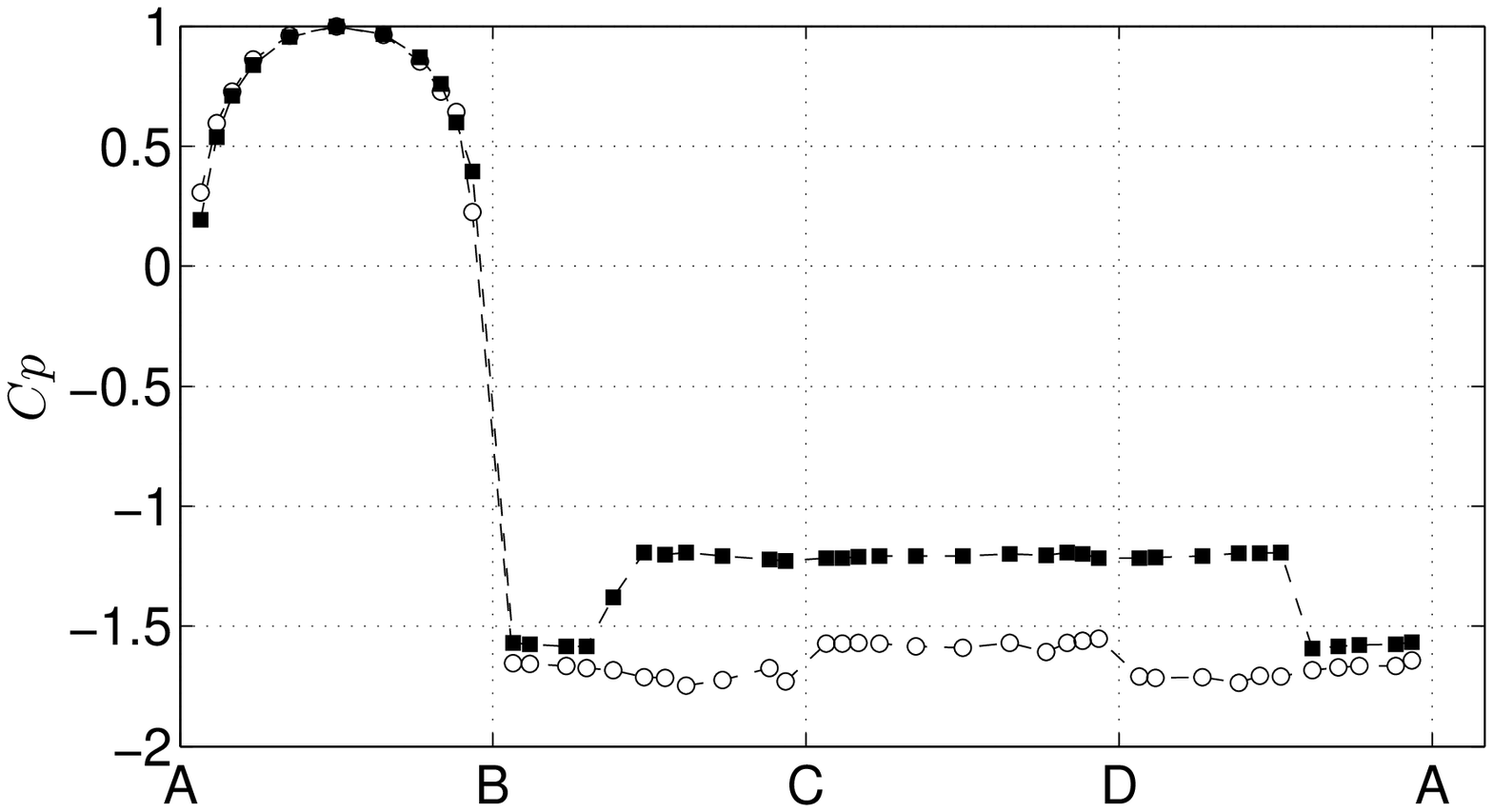}
\label{fig:CpRe40000}}
\hspace{0.2cm}
\subfigure[]
{\includegraphics[width=0.6\textwidth]{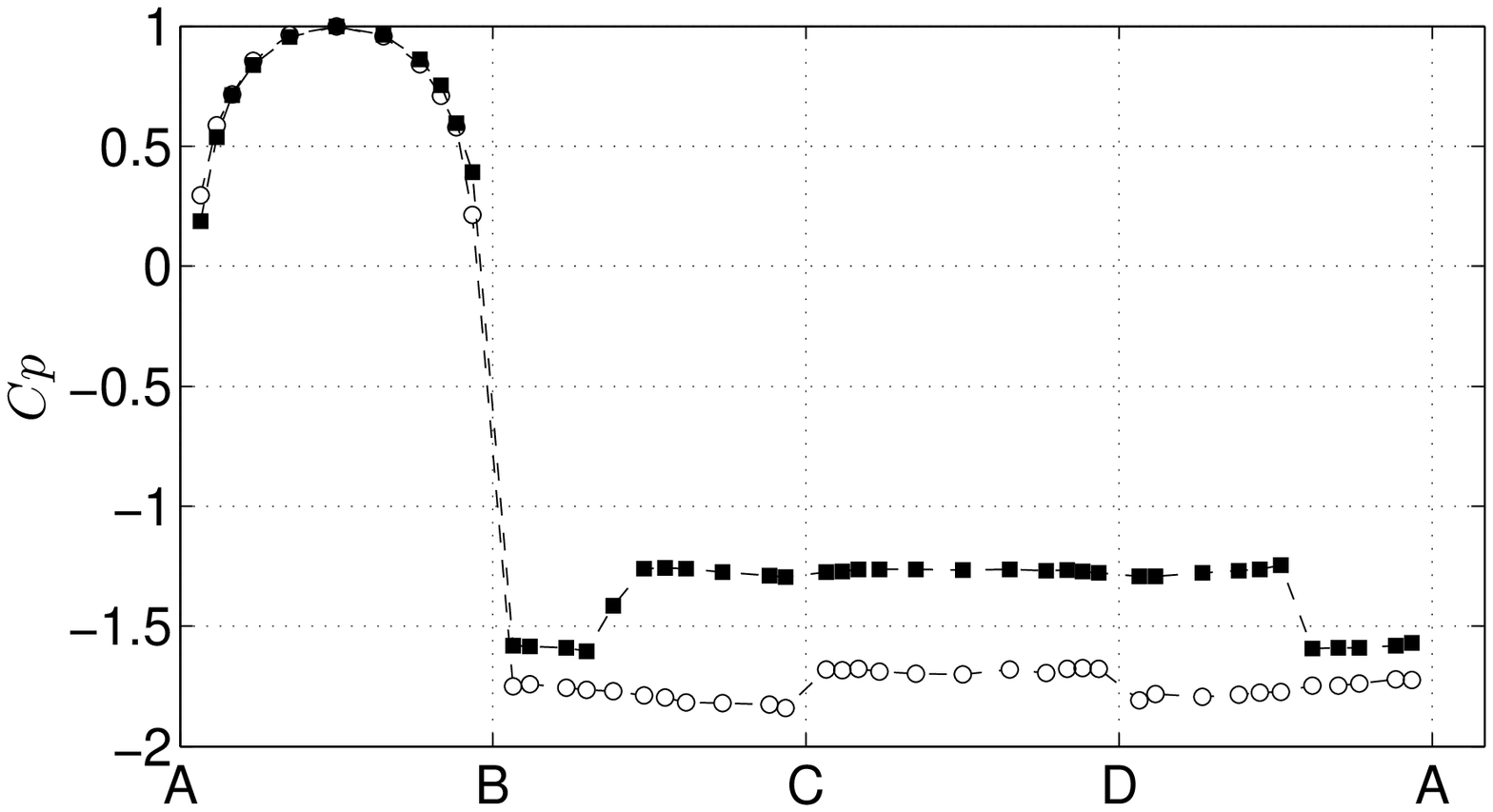}
\label{fig:CpRe60000}}
\caption{Pressure distribution around both the uncontrolled (open circles) and
the controlled (closed squares) square cylinders obtained for (a) $Re = 2 \cdot 10^4$, (b)
$Re = 4 \cdot 10^4$ and (c) $Re = 6 \cdot 10^4$.}
\end{figure}

In order to assess that simple mechanisms, the pressure distribution around the controlled
cylinder is compared to that obtained for the uncontrolled one as shown in Figs.
\ref{fig:CpRe20000}-\ref{fig:CpRe60000} for several $Re$.

\begin{figure}[htbp]
\centering
\includegraphics[width=0.8\textwidth]{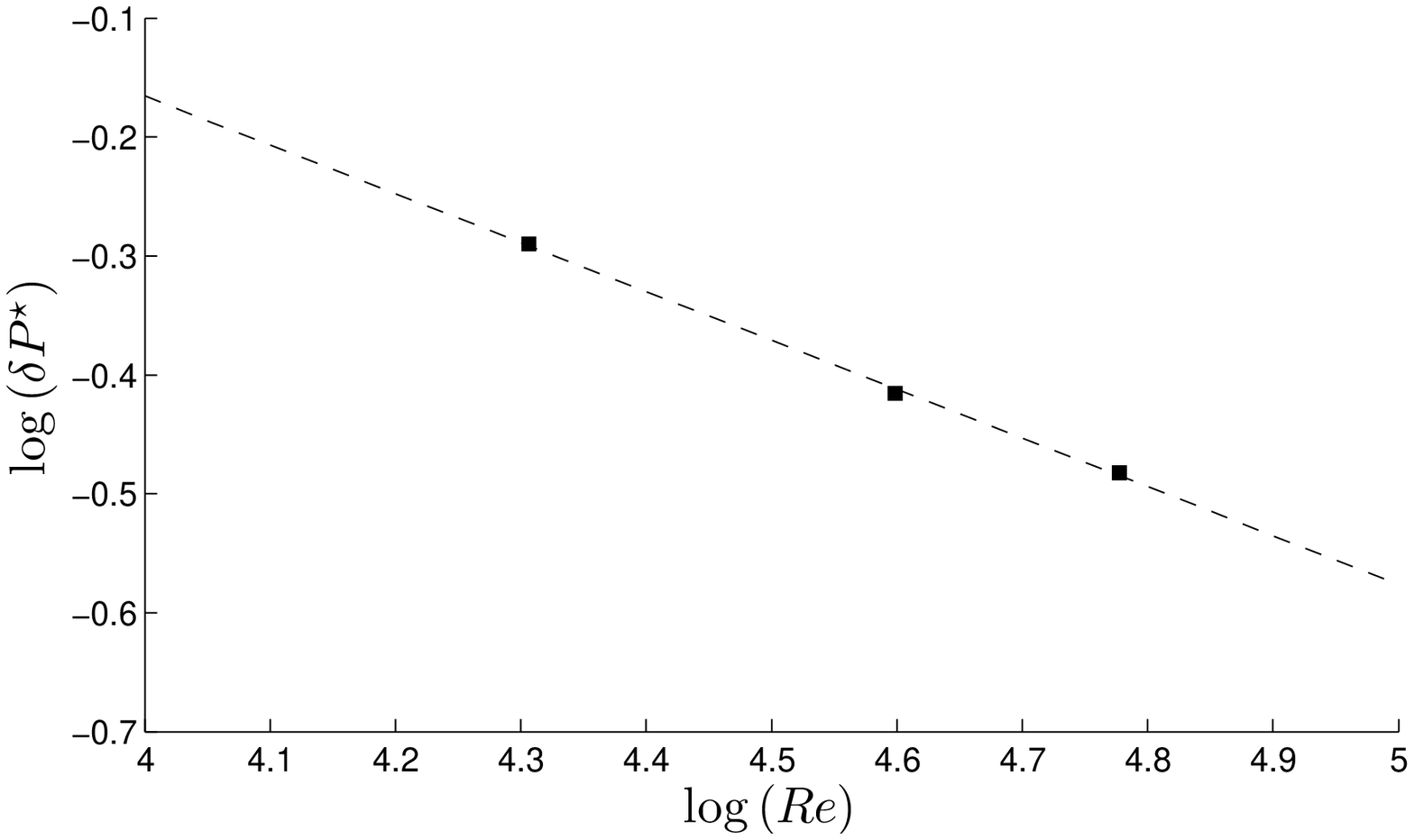}
\caption{Normalized pressure drop across the movable flaps as a function of $Re$ in log-log
representation. The dashed line symbolizes the curve $\delta P^\star = K Re^{-\gamma}$ with
$K \approx 32$ and $\gamma \approx 0.4$.}
\label{fig:dp}
\end{figure}

One can see that, independently of $Re$, the control system significantly modifies
the pressure distribution on the lateral sides (i.e. BC and DA) and the base (i.e.CD)
as well, whereas the stagnation face (i.e. AB) is unaffected. The pressure jump
$\delta p$ visible on the both sides of the controlled cylinder coincides with the
position of the leading edge of the movable flaps. Assuming that each flap can be
assimilated to a screen with a given blockage ratio and that $\delta P$ is
representative of the pressure drop across the flaps, it comes \cite{LawsLivesey78} 

\begin{equation}
\delta P^\star = \frac{\delta P}{\frac{1}{2} \rho U_\infty^2} \sim G(Re),
\end{equation}

where $\delta P^\star$ is the dimensionless pressure jump and $G(Re)$ a dimensionless
function. The evolution of $\delta P^\star$ computed from the pressure distribution
is plotted in Fig. \ref{fig:dp} with respect to $Re$. 

This plot points out that the normalized pressure drop $\delta p^\star$ decreases with
increasing $Re$ in agreement with the results reported by Groth and Johansson
\cite{GrothJohansson88}. In the present study, the fitting (in the least-mean square
sense) of our experimental data yields

\begin{equation}
\delta P^\star = K Re^{-\gamma},
\end{equation}

where $K \approx 32$ and $\gamma \approx 0.4$, at least for the range of $Re$
tested here. According to this result, the pressure drop $\delta P$ evolves like $Re^{1.6}$.
However, one can notice that the pressure upstream from the flap (left-hand side in Fig.
\ref{fig:WF}) remains almost constant with respect to $Re$. This confirms that the pressure
downstream from the flap (right-hand side in Fig. \ref{fig:WF}) increases with increasing
$Re$. However, we stress that care should be taken in extrapolating this trend
to higher $Re$. More data are required for a conclusive assessment of this issue
which is therefore left for future study.

\begin{figure}[htbp]
\centering
\includegraphics[width=0.7\textwidth]{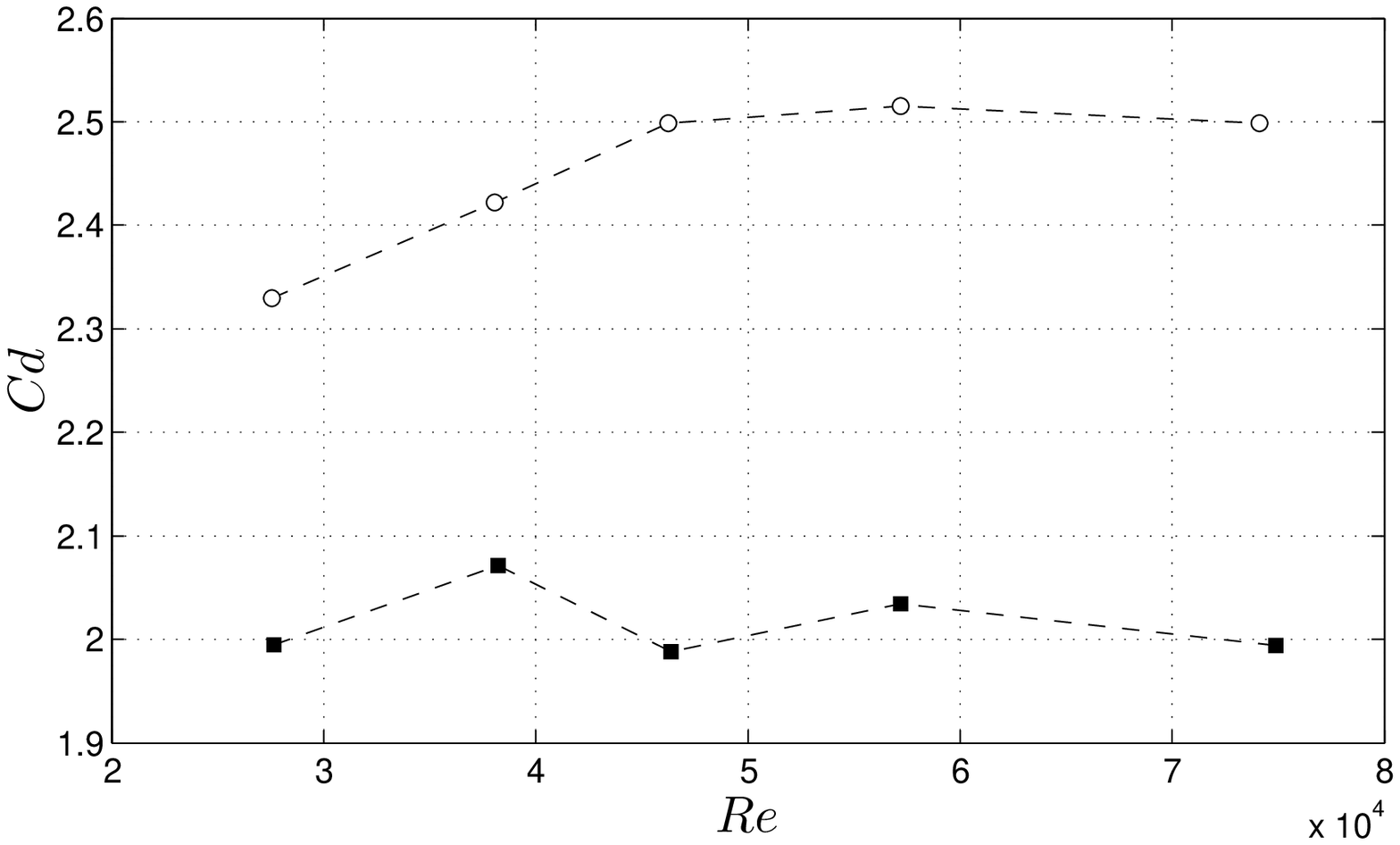}
\caption{Drag coefficient as a function of the Reynolds number measured
for the uncontrolled square cylinder (open circles) and for the controlled
square cylinder (close squares).}
\label{fig:Cd}
\end{figure}

The increase of the pressure level on the base of the controlled cylinder results from
the downstream displacement of the vortex core $V_2$ as illustrated in Fig. \ref{fig:WF}.
The main consequence of this phenomenon is the reduction of the drag force. In order to
evaluate the efficiency of the control system, the drag coefficient $Cd$ ($\equiv
\frac{D}{\rho U_\infty^2 H L / 2}$ with $D$ the drag force) has been measured by means of 
static load balance in both the uncontrolled and the controlled configurations. Fig.
\ref{fig:Cd} shows the evolution of $Cd$ as a function of $Re$. It is worth noticing that the values reported in this
study are not corrected for the blockage effect (see e.g. \cite{Maskell63}). However, we have
checked that the differences between the uncontrolled and the controlled cases are not affected
by these corrections. This can be
explained by the fact that the control system does not change the blockage ratio as
evidenced by the velocity profiles shown hereinbefore.

Fig. \ref{fig:Cd} evidences that the control system developed in this study improves
significantly the aerodynamics of the square cylinder. Indeed, the drag reduction equals
about 22\% in average over the entire range of $Re$ used in this study. This value is
comparable to that reported by Shao and Wei \cite{ShaoWei08} for similar $Re$. Furthermore,
these results confirm that the control seems more efficient with increasing $Re$.

\section{Conclusions}

An original passive control system consisting in a couple of porous flaps has been
developed and fitted on a square cylinder in such a way that the flaps can freely
rotate around their leading edge. These flaps have been designed to mimic the main
features of bird's feathers, i.e. the stiffness of the shaft coupled with the porosity
of the vane. This goal has been achieved by assembling a rigid and light plastic
skeleton with a porous fabric.

The device developed here can be assimilated to a self-adaptive passive control
system as it is activated by the vortex induced by the flow separation arising on
the side of the square cylinder. Due to a suction effect, the flaps depart from
the cylinder walls and flutter around a mean position. The dynamics of the movable
flaps has been investigated with respect to the Reynolds number. It has been shown
that the amplitude of the flap motion increases with increasing Reynolds number.
Furthermore, for low Reynolds number the rotation frequency of both flaps is
slightly higher than that of the natural vortex shedding even though their motion
seems asynchronous. However, the study of the relative motion of both flaps has revealed
that for high enough Reynolds number the flaps rotate in phase at a frequency
almost twice as small than the natural vortex shedding frequency. The spectral analysis
performed in the near wake of the cylinder has exhibited a lock-in coupling between the flow and the
movable flaps.

The efficiency of the control system has been evaluated by investigating the drag
force. Comparing both the uncontrolled and the controlled cases, about 22\% drag
reduction has been obtained over the entire range of Reynolds numbers tested in
this study. The investigation of the pressure distribution measured around the
square cylinders have evidenced that this enhancement of the aerodynamic performances
can be accounted for the increase of the pressure level on the base of the controlled
cylinder. A simple mechanisms accounting for this phenomenon has been suggested. In
this scenario, the pressure drop across the porous fabric induces an increase of
the pressure level downstream the flaps. As a consequence, the vortex core of the
recirculation region in the wake is expected to move downstream resulting in
the increase of the base pressure. This scenario is supported by the evolution
of the mean velocity reported in the wake.

Even though the results presented in this study are encouraging, further works are
required to get a better understanding of the control system and the flow. For that
purpose, the influence of relevant parameters such as the fabric porosity, for
instance, will be investigated in future.



\section*{Acknowledgements}
The authors are grateful to Mr. S. Loyer for his assistance concerning the experiments.

\end{document}